\documentclass[useAMS,usenatbib]{mn2e}
\usepackage{color,url,graphicx}
\usepackage{epstopdf}
\title[Distinguishing X-ray Processes Using the Fundamental Plane]{Using the Fundamental Plane of Black Hole Activity to Distinguish X-ray  Processes from Weakly Accreting Black Holes}
\author[Richard M. Plotkin et al.]%{%
{\parbox{\textwidth}{%
Richard~M.~Plotkin,$^{1}$\thanks{E-mail: r.m.plotkin@uva.nl} %; otheremail@otheraddress (ANO)} %
Sera~Markoff,$^{1}$ %
Brandon~C.~Kelly,$^{2}$\thanks{Hubble Fellow} %
Elmar K\"{o}rding,$^{3}$ \\
 and Scott~F.~Anderson$^{4}$} %
 \vspace{0.4cm} \\
	$^{1}$Astronomical Institute `Anton Pannekoek', University of Amsterdam, Science Park 904, 1098 XH, Amsterdam, the Netherlands
\\
	$^{2}$Harvard-Smithsonian Center for Astrophysics, 60 Garden St., Cambridge, MA 02138, USA\\
	$^{3}$Astronomy Department, University of Nijmegen, Postbus 9010, 6500 GL Nijmegen, the Netherlands \\
	$^{4}$Department of Astronomy, University of Washington, Box 351580, Seattle, WA 98195, USA 
	}

\newcommand{\aj}{AJ}
\newcommand{\mnras}{MNRAS}
\newcommand{\apjs}{ApJS}
\newcommand{\apj}{ApJ}
\newcommand{\apjl}{ApJL}
\newcommand{\araa}{AR\&A}
\newcommand{\apss}{Ap\&SS}
\newcommand{\aap}{A\&A}
\newcommand{\iaucirc}{IAU~Cric.}
\newcommand{\nat}{Nature}   %Nature
\newcommand{\aaps}{A\&AS}
\newcommand{\bl}{BL~Lac}
\newcommand{\mbh}{M_{\rm BH}}
\newcommand{\msun}{{\rm M_{\sun}}}
\newcommand\ion[2]{#1$\;${\scshape{#2}}}%                       % ion, i.e., CII = \ion{C}{ii}
\newcommand{\smbh}{SMBH}
\newcommand{\nblmass}{55}   % Number of SDSS BL Lacs I fit onto the plane

\defcitealias{merloni03}{M03}
\defcitealias{falcke04}{F04}
\defcitealias{koerding06}{K06}
\defcitealias{kelly07}{K07}
\defcitealias{plotkin10}{P10}

\begin{document}

\date{}

\pagerange{\pageref{firstpage}--\pageref{lastpage}} \pubyear{2011}

\maketitle

\label{firstpage}

%-----------------------------------------------------------------------------------------------------------------------------

\begin{abstract}
The fundamental plane of black hole activity is a relation between X-ray luminosity, radio luminosity, and black hole mass for hard state Galactic black holes and their supermassive analogs.  The fundamental plane suggests that, at low-accretion rates, the physical processes regulating the conversion of an accretion flow into radiative energy could be universal across the entire black hole mass scale.    However, there is still a need to further refine the fundamental plane in order to better discern the radiative processes and their geometry very close to the black hole, in particular the source of hard X-rays. Further refinement is necessary because error bars on the best-fit slopes of the fundamental plane  are generally large, and also the inferred coefficients can be sensitive to the adopted sample of black holes.   In this work, we regress the fundamental plane with a  Bayesian technique.  Our approach shows that sub-Eddington  black holes emit  X-ray emission that is predominantly optically thin synchrotron radiation from the jet, provided that their radio spectra are flat or inverted.  X-ray emission dominated by very radiatively inefficient  accretion flows are excluded at the $>$3$\sigma$ level.   We also show that it is difficult to place FR~I galaxies onto the fundamental plane because their X-ray jet emission is highly affected by synchrotron cooling.   On the other hand, \bl\ objects (i.e., relativistically beamed sub-Eddington AGN) fit onto the fundamental plane.   Including a uniform subset  of high-energy peaked BL~Lac objects from the Sloan Digital Sky Survey, we find  sub-Eddington black holes with flat/inverted radio spectra follow $\log L_x  = (1.45 \pm 0.04) \log L_R - (0.88 \pm 0.06) \log \mbh  - 6.07 \pm 1.10$, with $\sigma_{int}=0.07 \pm 0.05$~dex.  Finally, we discuss how  the effects  of synchrotron cooling of jet emission from the highest black hole masses can  bias fundamental plane regressions, perhaps leading to incorrect inferences on  X-ray radiation mechanisms.
\end{abstract}

%-----------------------------------------------------------------------------------s------------------------------------------


%-----------------------------------------------------------------------------------------------------------------------------
\section{Introduction}
 Accreting supermassive black holes (\smbh s), i.e., Active Galactic Nuclei (AGN), are thought to have important cosmological ramifications.  The immense energy output by an AGN can feedback with the environment near a SMBH and potentially influence the formation of galaxies \citep[e.g.,][]{dimatteo05, croton06, fontanot06, hardcastle07}.   Thus, accreting black holes are more than interesting laboratories to explore physics around extreme gravitational potentials.  They are also critical components for any self consistent view of galaxy formation.  However, the underlying physics of black hole accretion, and especially the mechanisms connecting the accretion flow with large scale outflows (i.e., the so-called disk/jet connection), are still outstanding problems in astrophysics. 

An increasing number of studies over the past few years suggest that accretion physics scales globally across the entire black hole mass scale, from $\sim$10~${\msun}$  Galactic black holes (GBHs) to 10$^6$--10$^{10}$~$\msun$ \smbh s  \citep[e.g.,][]{done05, jester05, nipoti05, koerding06_agnmapping, mchardy06, koerding07, markoff08, gliozzi10, kelly11}.  The discovery of the fundamental plane of black hole activity (hereafter the FP) -- a relation between X-ray luminosity ($L_X$), radio luminosity ($L_R$), and black hole mass ($\mbh$) for low-accretion rate black holes -- is one of the most prominent pieces of evidence supporting the unification of GBHs and \smbh s \citep[][hereafter \citetalias{merloni03} and \citetalias{falcke04}, respectively]{merloni03,falcke04}.    That is, once normalized by black hole mass, the process that regulates the fraction of energy accreted onto a black hole that is ultimately converted into radiation could be universal for weakly accreting black holes.  The FP  is a natural consequence if black hole accretion and relativistic jet physics are scale invariant.  In other words, any spatial dependence of physical properties like magnetic field strength, jet power, etc., can be scaled to a characteristic size (e.g., the gravitational radius $r_g=GM/c^2$) for all black holes \citep{falcke95, heinz03, markoff03}.    We define the FP in this work as
\begin{equation}
\log L_X = \xi_R \log L_R + \xi_M \log M_{BH} + B,
\label{eq:fp}
\end{equation}
and we refer to $\xi_R$ and $\xi_M$ as the FP coefficients.

The FP implies that, under certain conditions, it is reasonable to apply our general knowledge of GBHs toward AGN, and vice versa.  Opposing ends of the mass scale offer complementary insight, and the FP therefore allows a more holistic view of black hole accretion and jet physics \citep[see, e.g.,][]{markoff10}.   For example, GBHs present the opportunity to follow dramatic changes in accretion processes in real time: GBH outbursts and subsequent dimming back to quiescence  typically last only months to years.  During an outburst, we often witness the launching and quenching of a radio jet.   GBHs also undergo state transitions marked by different contributions of hard non-thermal  X-ray emission (from a jet/corona) to the total observed X-ray flux, where softer X-rays are emitted by the accretion disk \citep[see, e.g.,][]{fender04,mcclintock04, homan05, fender09}.  

The ``human timescale'' is a major advantage that stellar mass black holes hold over \smbh s, since any equivalent AGN episode would last 10$^7$--10$^8$ times longer.  On the other hand, we only know of around two to three dozen GBHs in our Galaxy \citep[see][]{dunn10}, precluding large statistical studies, and we cannot yet predict when GBHs will undergo outbursts.    AGN are much more common, with over a million so far catalogued \citep{richards09}, offering the opportunity for statistical studies to obtain more robust constraints on general trends.   AGN central black hole masses  also cover almost four decades in mass, thus providing a larger dynamic range than  GBHs when  searching for empirical constraints on black hole mass scalings.  

The FP can be exploited to distinguish the radiative processes responsible for emission originating very close to an accreting black hole.  The nature of this emission especially in the X-ray waveband  is still a controversial subject \citep[see, e.g., \S 3.2 of][]{narayan05}.   Observed X-rays are likely a superposition of several components, potentially including synchrotron radiation from a jet, synchrotron self Compton (SSC), emission from the accretion flow, and inverse Compton scattering of lower energy photons off a corona.  All of these components likely contribute to the observed X-ray spectra at some level.  However, we may expect a specific component(s) to dominate under certain conditions (see, e.g.,  \citealt{markoff01}; \citealt{markoff03}; \citealt{markoff05}; also see \citealt{gilfanov10} for a recent review on X-ray emission  from X-ray binary systems).    \citetalias{merloni03} show that different FP coefficients are predicted in the case that X-rays are predominantly optically thin synchrotron jet emission, compared to the scenario where X-rays originate by inverse Compton scattering.   The slope of the plane can further diagnose the radiative efficiency of the accretion flow.    Several studies over the past few years  (including the original discovery papers) have indeed used the FP to investigate the origin of X-rays, but with varied success (\citealt{koerding06}, hereafter \citetalias{koerding06}; \citealt{wang06, li08, gultekin09, yuan09}). 

A major challenge to realizing the plane's full potential to diagnose X-ray emission is that large (non-isotropic) observational uncertainties are associated with $L_X$, $L_R$, and especially $\mbh$.  Thus, inferring reliable coefficients  becomes a challenging multivariate regression problem requiring advanced statistical methods.  \citetalias{koerding06} investigate this issue by exploring different regression techniques while incorporating detailed error budgets for multiple samples of accreting black holes.  They show that improper accounting of observational error bars will bias the best-fit values of $\xi_R$ and $\xi_M$.    

\citetalias{koerding06} also show that the inferred coefficients and the intrinsic scatter around the FP depend on the sample used to fit the plane.   A  global radio/X-ray correlation is observed for quiescent GBHs in the hard (i.e., low-accretion rate, non-thermal)  state \citep{corbel03, gallo03}.  When more luminous (soft-state, thermally dominated) GBHs are included with higher accretion-rates, the radio jet is quenched and the correlation breaks down  \citep{fender99, gallo03}.  The FP is essentially an extension of the hard-state GBH radio/X-ray correlation to include a mass term.   \citetalias{koerding06} show that the intrinsic scatter about the FP is smallest when only considering GBHs and AGN with the lowest  Eddington normalized luminosities.   When adding higher-luminosity sources, the FP  coefficients change and the intrinsic scatter increases.   The apparent sensitivity of the FP on accretion rate supports the notion that AGN display accretion ``states'' similar to GBHs \citep{meier01, maccarone03, fender04, jester05, markowitz05, koerding06_agnmapping, koerding07}.

 \citet{gultekin09} further explored the sensitivity of the FP regression on the adopted black hole sample.    Previous FP studies used highly heterogeneous samples, particularly for the AGN, assembled from different surveys using different telescopes.   \citet{gultekin09} overcame such problems by using a (quasi)-volume-limited sample of AGN, with all X-ray data obtained by \textit{Chandra} and analyzed uniformly.  They also only consider AGN with dynamical black hole mass estimates from \citet{gultekin09_msigma}, which are more reliable than secondary scalings like reverberation mapping  \citep[see, e.g.,][]{onken04, peterson04}.   The work of \citet{gultekin09} is a powerful proof of concept: they show a carefully-crafted uniform sample has the potential to sharpen the plane enough to eventually  yield physically meaningful constraints on black hole accretion physics and unification.  However, even though their AGN sample covers a large range of black hole masses and accretion rates, it only includes 18 AGN.  This is much too small to distinguish radiation mechanisms.  Interestingly, even with a small sample, they already confirm that  the inclusion of high-accretion rates sources (especially Seyfert galaxies) increases the intrinsic scatter about the FP.

There is an overwhelming need to further refine the FP.  We must have more confidence in the derived best-fit coefficients in order to place more meaningful constraints on the physical processes connecting accretion inflows and outflows.  More reliable coefficients are also required if the FP is to ever become a viable method to accurately estimate black hole masses (assuming one has radio and X-ray luminosity measurements).  Here, we focus on sub-Eddington AGN ($\dot{m} < 0.01-0.02$) that we expect to be in an analogous spectral state to hard state GBHs.  Since these objects display a narrower intrinsic scatter about the plane, such a sample will allow a more rigorous exploration on how statistics and sample selection might bias the inferred slopes of the FP.  

A brief summary of the theoretical derivation of the FP is outlined in \S \ref{sec:fptheory}.  Then, for the first time, we fit the FP with a Bayesian multivariate regression technique.  We find that our adopted statistical approach can more tightly constrain X-ray radiative processes than previous studies (\S \ref{sec:fpfit}).  Then, in the context that X-rays are predominantly optically thin synchrotron jet emission, we explore the effects of sample selection by attempting to place a large uniform sample of \bl\ objects on the FP (\S \ref{sec:fitBL}).  We consider \bl\ objects because, due to beaming effects,  we know that their spectral energy distributions (SEDs) are dominated by jet emission (even in the X-ray).  Also, they host the most massive black holes ($\mbh\sim10^8-10^9~\msun$), so they provide the largest possible dynamic range when compared to GBHs.   Finally, we compare to previous FP studies in \S \ref{sec:disc}, and we summarize our results in \S \ref{sec:summary}.  Throughout we define spectral indices as $f_{\nu} \sim \nu^{-\alpha}$; we use $\dot{M}$ to express accretion rates in physical units, and normalized (i.e., unitless)  accretion rates are written as $\dot{m}\equiv \dot{M}/\dot{M}_{Edd}$.  We adopt a standard cosmology: $H_0=71$~km~s$^{-1}$~Mpc$^{-1}$; $\Omega_m=0.27$; and $\Omega_\Lambda=0.73$.  

\section{Deriving the Plane: Theoretical Scalings of Multiwavelength Emission with Black Hole Mass}
\label{sec:fptheory}
The FP was independently discovered by  \citetalias{merloni03} and \citetalias{falcke04}, who derived the expected coefficients $\xi_R$, $\xi_M$, and $B$ (Equation~\ref{eq:fp}) under the assumption that the radio emission is synchrotron from a scale-invariant conical jet.\footnote{The mathematical representation for a scale invariant jet states that any quantity $f$ required to describe the jet structure (e.g., magnetic field strength) can be independently separated into a normalization set only by boundary conditions, $\phi_f$, and a structure function describing how the jet changes with scaled radius, $\psi_f(r/r_g)$: $f(r,M,\dot{m}) = \phi_f(M,\dot{m}) \psi_f(r/r_g)$  \citep{heinz03}.}  %%
We refer the reader to the discovery papers for details.  Briefly, the amount of synchrotron radiation emitted from a scale invariant  jet depends on the black hole mass and accretion rate \citep{falcke95,heinz03}.  Substituting the appropriate expression for accretion rate as a function of  X-ray luminosity (which can also depend on black hole mass)  into the expression for radio luminosity leads to the FP.    The exact expression to use for X-ray luminosity depends on if the X-rays are modeled predominantly as optically thin jet synchrotron or  inverse Compton emission, and different coefficients are predicted in each case.

 We stress that   jet synchrotron and inverse Compton  emission  are not  mutually exclusive.  Even if a jet's optically thin synchrotron radiation extends into the X-ray, then the underlying accretion flow  can also still produce X-rays.   Throughout this text we will make comparisons to the theoretical predictions in \citetalias{merloni03} and \citetalias{falcke04}, and we will make statements on whether the X-rays are emitted primarily by the `jet' or by the `corona.'  The terms `jet' and `corona' are meant only to designate the mechanism emitting \textit{most} of the X-rays (i.e., optically thin jet synchrotron from a magnetized collimated outflow vs.\ inverse Compton off a corona, respectively), and should not be misinterpreted to mean only a jet or only a corona is present.   In fact, the presence of multiple components contributing to the observed X-rays likely contributes to the observed intrinsic scatter about the FP.
   
The results of the two FP discovery papers are consistent with each other, and the two papers generally make similar assumptions.  For example, both papers assume that effects due to different black hole spins from source to source are of second order and will be manifested only by introducing excess scatter in the FP.     Both papers also assume that radiative losses from synchrotron cooling are not important, because radiative cooling affects the emitting particles' dynamics.    The latter is an important assumption, as the predicted FP coefficients for the `jet' model are only applicable if X-rays are strictly optically thin (i.e., uncooled) synchrotron.  A derivation of the FP that includes synchrotron radiative cooling losses is presented in \citet{heinz04}.   

 Despite the above similarities, the two groups took  different approaches.   For example, \citetalias{falcke04} derive the FP specifically for the case of sub-Eddington black holes assuming X-rays are  optically thin synchrotron  emission from jets.   \citetalias{merloni03} consider all accretion rates, and they also derive more general expressions applicable not just to optically thin synchrotron X-rays, but also where X-rays are inverse Compton.    An important difference between the two discovery papers is they use different notation: \citetalias{merloni03} define $\log L_R$ as the dependent variable and fit $\log L_R = \xi_{RX}\log L_X + \xi_{RM}\log M + c$, compared to \citetalias{falcke04} who cast $\log L_X$ as the dependent variable.  We adopt \citetalias{falcke04}'s notation in this work\footnote{We use the notation of \citetalias{falcke04} here primarily because it was also used by \citetalias{koerding06}.  We will eventually compare coefficients to those derived by \citetalias{koerding06}, so this is the more natural notation for us.}  %%
 (see Equation~\ref{eq:fp}), which is related to \citetalias{merloni03}'s notation  by $\xi_R = 1/\xi_{RX}$ and $\xi_M = -\xi_{RM}/\xi_{RX}$.     In Sections~\ref{sec:jetcoeff} and \ref{sec:iccoeff} we transcribe \citetalias{merloni03}'s expressions for the FP coefficients (their Equations 11 and 13) to our notation for both the optically thin synchrotron jet and  inverse Compton models for clarity, as we will eventually compare our results to their theoretical predictions.  Note, the theoretical formulae for the  FP coefficients do not account for intrinsic scatter about the plane, so it is possible to convert between the two notations.  However, in reality there is intrinsic scatter, so one cannot regress a sample of black holes in one notation, and then convert the best-fit regression coefficients to the other notation using the above formulae.  Instead, the regression must be repeated adopting a different dependent variable.
 
Finally, we note that the FP has been criticized as a spurious correlation driven by a similar dependence of $L_R$ and $L_X$ on distance \citep{bregman05}.  We show in \S \ref{sec:fpfit}  that the FP coefficients for sub-Eddington black holes match those predicted by the jet model in \citetalias{merloni03} and \citetalias{falcke04}. On a  qualitative level, it would be a remarkable coincidence for a common dependence on  distance  to exactly produce  the predicted non-linear correlation.  Quantitatively, \citet{merloni06} show in great detail through multiple statistical tests that the FP is not simply an artifact of plotting distance vs.\ distance;  other authors  reach the same  conclusion (\citetalias{merloni03,koerding06}; \citealt{wang06}). 

\subsection{Expected FP Coefficients for Optically Thin Synchrotron X-rays}
\label{sec:jetcoeff}
  
 In the case that X-rays are dominated by optically thin synchrotron from a jet:
\begin{eqnarray}
\nonumber \left(\xi_{R}\right)_{jet} & = & \frac{(p+4)(p+5)}  {2(2p + 13 + \alpha_R p + 6\alpha_R)}, \\[0.2cm]
		  \left (\xi_M \right)_{jet} & = & 3 - \frac{(p+5) (2p + 13 + 2\alpha_R)}  {2(2p + 13 + \alpha_R p + 6\alpha_R)}
\label{eq:jetcoeff}
\end{eqnarray}
where p is the power law index of the accelerated relativistic electrons (i.e., $dn_e/d\gamma \sim \gamma^{-p}$), with values $p\sim2-3$ typical, and we can substitute $p=2\alpha_X+1$ for optically thin synchrotron, where $\alpha_X$ is the observed X-ray spectral index.  Note, our lack of knowledge about the physical conditions in the jet is absorbed into two observable quantities, the radio and  X-ray  spectral indices $\alpha_R$ and $\alpha_X$.   Since the radio and X-rays are emitted by the same underlying source, we expect  less intrinsic scatter about the FP if X-rays are optically thin jet synchrotron compared to inverse Compton.

\subsection{Expected FP Coefficients for Inverse Compton X-rays}
\label{sec:iccoeff}
For the case that the X-rays originate as inverse Compton from a corona:
\begin{eqnarray}
\nonumber \left(\xi_R\right)_{cor} & = & \left( \frac{\partial \ln \Phi_B}{\partial \ln \dot{m}}\right)^{-1}    \left( \frac{q(p+4)}   {2p+13+\alpha_Rp + 6\alpha_R}   \right), \\[0.2cm]
\nonumber \left(\xi_M\right)_{cor} & = & 1 - q  \left( \frac{\partial \ln \Phi_B}{\partial \ln \dot{m}} \right)^{-1} \left( \frac{\partial \ln \Phi_B}{\partial \ln M}\right)   \\[0.2cm]
		  	    &     &	-  \left(\frac{\partial \ln \Phi_B} {\partial \ln \dot{m}}\right)^{-1} \left(\frac{q(2p+13+2\alpha_R)}{2p+13+\alpha_R p + 6\alpha_R}\right)  
\label{eq:coronacoeff}
\end{eqnarray}
where $\Phi_B(M,\dot{m})$ describes the boundary conditions of the magnetic field at the base of the  jet, and $q$ parameterizes the radiative efficiency of the accretion flow ($L_X \sim M\dot{m}^q$).   Values  of $q$, $\partial \ln \Phi_B / \partial \ln M$, and $\partial \ln \Phi_B / \partial \ln \dot{m}$ for different radiation mechanisms can be found in Table~3 of \citetalias{merloni03}.    

\subsubsection{Radiatively Inefficient Accretion Flows}
\label{sec:riaf}
Sub-Eddington black holes are most likely fed by radiatively inefficient accretion flows (RIAFs), so we will primarily consider values of $q\geq2$ when comparing FP regressions to theoretical predictions (radiatively efficient accretion disks have $q=1$).  Any such mechanically cooled accretion flow should have $\Phi_B^2\propto M^{-1} \dot{m}$ (\citealt{heinz03}; \citetalias{merloni03}).  Following \citetalias{merloni03}, the specific RIAF model we will consider here is an advection dominated accretion flow \citep[ADAF;][]{narayan94, abramowicz97}, fixing the viscosity parameter to $\alpha_{\rm v}=0.1$, the ratio of gas to magnetic pressure to $\beta=10$, and the fraction of turbulent energy in the plasma that heats the electrons to $\delta=0.3$.  The precise value of $q$ for a particular accreting black hole depends on $\dot{m}$ (see Equation 12 of \citetalias{merloni03}).  \citetalias{merloni03} show that, assuming the above parameters for an ADAF, sub-Eddington black holes with $\dot{m} < 2\times10^{-2}$ should have an average $q=2.3$.   Thus, in this work, we will use $q=2.3$, $\partial \ln \Phi_B / \partial \ln M=-0.5$, and $\partial \ln \Phi_B / \partial \ln \dot{m}=0.5$  to compare to \citetalias{merloni03}'s theoretical FP coefficients for inverse Compton dominated X-rays.   Differences in $q$ between objects (as well as the likely possibility that $q$ varies with time) will contribute to some of the observed scatter about the FP.  

As noted in \citetalias{merloni03}, the above ADAF model represents a rather general description of a RIAF.  Since the FP is a statistical correlation,  FP regression coefficients  only provide   information on the \textit{average}  properties of statistical samples of accreting black holes.  In fact, every accreting black hole will have its own `FP coefficients' that can vary with time, depending on its specific values of $\alpha_R$, $p$, and $q$.  Furthermore, an individual black hole at a given accretion rate can even have different fractions of optically thin jet vs.\ inverse Compton emission contributing to the total observed X-rays depending on if it is heading into outburst or declining back into quiescence \citep[e.g.,][]{russell10}.  We thus aim to keep our accretion model as general as possible so we can focus primarily on average trends.  

There are more complicated RIAF models in the literature.   For example, RIAFs can have convective instabilities if $\alpha_{\rm v}$ becomes small enough  (i.e., so-called convection dominated accretion flows or CDAFs; e.g, \citealt{narayan00}).  RIAFs can also produce  strong mechanical outflows (i.e., advection dominated inflow-outflow solutions or ADIOS; e.g., \citealt{blandford99}).   In such alternatives  to ADAFs,  X-ray emission is likely dominated by thermal bremsstrahlung\footnote{We note, however, rapid X-ray variability (as is often observed) would usually rule out thermal bremsstrahlung.}, %%
which has $q=2$ (and $\Phi_B^2\propto M^{-1}\dot{m}$, like a RIAF).  For example, \citet{yuan02}  show that  only for unphysical values of $\delta=1$ (i.e., all the viscous dissipation goes toward heating the electrons) can ADIOS models predict that a radiative process other than  thermal  bremsstrahlung  will dominate the observed X-ray spectrum of Sgr~A$^{\star}$ in the quiescent state.   To capture the basic general characteristics of alternative RIAF models, we will also explicitly  compare FP regressions to theoretical predictions assuming $q=2.0$.   We  note though,  according to Equation~12 of \citetalias{merloni03}, values of $q\sim2$ could also be consistent with the pure ADAF model described above (and used by \citetalias{merloni03}) if one does not consider any black holes with $\dot{m}< 10^{-4}$.    Throughout this paper, we will therefore use the more general term RIAF to describe inefficient accretion, and differences in RIAF models will be parameterized in a very general sense by the radiative efficiency $q$.

\begin{figure}
\includegraphics[scale=1.0]{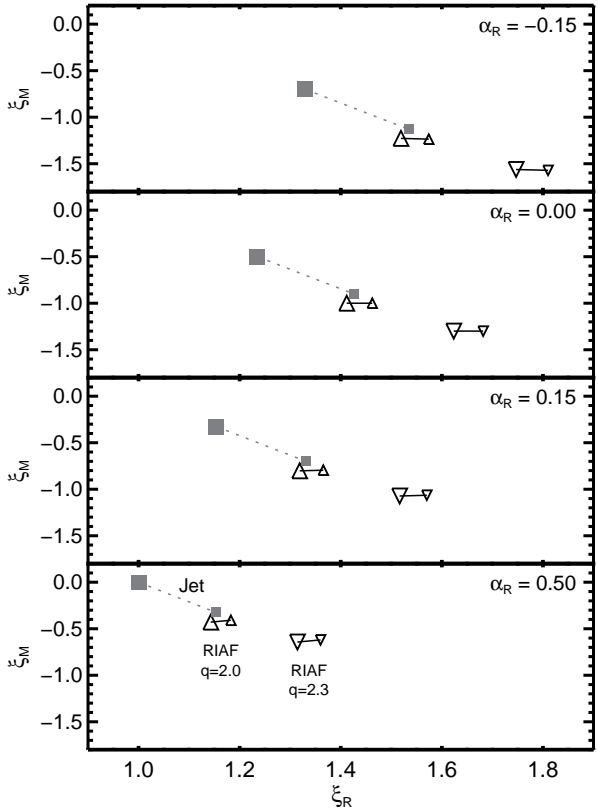}
 \caption{Expected black hole mass ($\xi_M$) vs.\ radio luminosity ($\xi_R$) coefficients from Equations~\ref{eq:jetcoeff} and \ref{eq:coronacoeff}, in the cases that X-rays are dominated by optically thin jet synchrotron (filled squares) or by inverse Compton off an X-ray corona (open symbols).  For the corona model, we show predictions from a radiatively inefficient accretion flow (labelled as RIAF)  with radiative efficiencies $q=2$ (triangles) and 2.3 (upside-down triangles).   Each model is illustrated by a pair of points connected with a dashed or solid line for the jet or RIAF models, respectively.   These sets of lines/points show the range of expected coefficients for  typical values of $p$, ranging from $p=2$ (larger data points) to  2.8 (smaller data points).  The top panel is for a radio jet with an inverted radio spectrum ($\alpha_R=-0.15$), with the radio spectral index becoming increasingly steeper moving downwards. 
  }
 \label{fig:coeff}
 \end{figure}

 The expected coefficients for  jet synchrotron (Equation~\ref{eq:jetcoeff}) vs.\ inverse Compton (Equation~\ref{eq:coronacoeff}) dominated X-rays are shown in Figure~\ref{fig:coeff} (for different values of $\alpha_R$, $p$, and $q$).  Although different emission processes generally predict different coefficients,  the solutions are also mildly degenerate in that some combinations of models, $\alpha_R$, $p$, and $q$ can predict similar coefficients.    For illustrative purposes, we  explicitly show predicted FP coefficients for RIAF/corona models assuming $q=2.0$ and 2.3.   These two values represent (approximate) extremes of a realistic range of plausible radiative efficiencies, as explained above.  However,  other values of $q$ are also possible.

\section{Fitting the Plane: Are FP Slopes Biased by Statistical Effects?}
\label{sec:fpfit}
 The observed multivariate correlation between multiwavelength emission and black hole mass   supports the idea of black hole mass scaling, but the mere presence of a correlation on its own does not provide insight into the physics governing black hole accretion.  Instead, one must compare the slope of the FP to  predictions from physical models.   Therefore,  care must be taken when regressing the FP, to ensure the inferred slopes  are not highly impacted by statistical effects.  In this section, we perform a Bayesian regression analysis on a sample of accreting black holes, and we compare to the Merit function, which is the regression technique most commonly used by previous FP studies.

\subsection{Sample of Low-accretion Rate Black Holes from \citetalias{koerding06}} 
\label{sec:kfc}
As stated earlier, low-accretion rate black holes across the mass scale are expected to be in a similar ``hard-like'' accretion state, and they show the least scatter about the FP.   Considering only low-accretion rate objects will therefore allow more sensitive constraints on  how statistics may influence the FP.  Here, we use a sample of low-accretion rate black holes assembled by \citetalias{koerding06}, which, following \citetalias{koerding06},  we refer to as the KFC sample.\footnote{The KFC sample is based on the sample compiled by \citetalias{falcke04}, and it is modified to include more LLAGN and to exclude Seyfert galaxies. }   %%

     The KFC sample includes radio luminosities, $L_R=\left(\nu L_{\nu}\right)_{5GHz}$, X-ray luminosities in the 0.5-10~keV band, and black hole masses for 77 accreting black holes.  Included are  25 observations of GBHs  in the hard state (from GX 339-4, V404 Cyg, 4U 1543-47, XTE 1118+480, and XTE J1550-564).  There are 52 AGN, including one observation of Sgr~A$^\star$ ($\sim$10$^6~\msun$) during a hard X-ray flare, 17 low-luminosity AGN (LLAGN; $\sim$10$^7$--10$^8$~$\msun$), and 15 FR~I galaxies and 19 \bl\ objects ($\sim$10$^8$--10$^9$~$\msun$).    We use the KFC sample because it contains exclusively low-accretion rate black holes, and its regression with various techniques has already been explored  by \citetalias{koerding06}.    We exclude Seyfert galaxies, since their inclusion would increase the intrinsic scatter, and they may not be supermassive analogs to hard state GBHs.  We refer the reader to \citet{panessa07} for more information on Seyferts regarding the FP.
  
  \citetalias{koerding06} subdivide their sample into various subsets to test the effect of sample selection on the inferred FP slopes.  We test the Bayesian regression on each of their subsets that includes GBHs and excludes Seyfert galaxies, leaving four KFC subsamples (see Table~\ref{tab:samples}) which we refer to  as the \textit{full sample} (GBHs, Sgr~A$^{\star}$, LLAGN, FR~Is, and \bl s; 77 objects), the \textit{contracted subsample} (GBHs, Sgr~A$^{\star}$, LLAGN; 43 objects), the \textit{contracted+\bl\  subsample} (GBHs, Sgr~A$^{\star}$, LLAGN, \bl s; 62 objects), and the \textit{contracted+FR~I subsample} (GBHs, Sgr~A$^{\star}$, LLAGN, FR~Is; 58 objects).  Although not discussed in this section, three other samples (called KFC+SDSS, KFC+SDSS-HBL, and KFC+SDSS-LBL) are introduced in \S\ref{sec:fitBL} and included in Table~\ref{tab:samples} for completeness.
    
 \begin{table}
\caption{Samples Used to Regress the FP}
\label{tab:samples}
\scriptsize
\begin{tabular}{l c l }
 \hline
		Sample      &  %1
	         Total           &  %2           
	         	Notes  	   \\   %3

\hline
 contracted                 &   43   & GBH, SgrA$^{\star}$, LLAGN \\
  contracted+\bl        &   62   &   GBH, SgrA$^{\star}$, LLAGN, \bl \\
 contracted+FR~I    &   58   &   GBH, SgrA$^{\star}$, LLAGN, FR~I \\
 full                               &  77   &  GBH, SgrA$^{\star}$, LLAGN, FR~I, \bl \\ \hline
KFC+SDSS            &   98   &   GBH, SgrA$^{\star}$, LLAGN, SDSS \bl \\
KFC+SDSS-HBL   &   82    &   GBH, SgrA$^{\star}$, LLAGN, SDSS HBLs \\
KFC+SDSS-LBL    &  59      &   GBH, SgrA$^{\star}$, LLAGN, SDSS LBLs \\
\hline
\end{tabular}

\medskip
The KFC+SDSS,  KFC+SDSS-HBL, and KFC+SDSS-LBL samples are described in \S\ref{sec:fitBL}.  The other four are defined in \S \ref{sec:kfc}.
\end{table}

 \subsubsection{Synchrotron Cooling}
 \label{sec:kfcSynchCooling}
 The derivation of the FP in both \citetalias{merloni03} and \citetalias{falcke04} assumes synchrotron cooling is unimportant.  The frequency where electrons undergo significant cooling is anti-correlated with black hole mass, and  \citetalias{falcke04} and \citetalias{koerding06} argue this frequency is below the X-ray band for the most massive black holes.  Synchrotron cooling is thus a concern for FR~I galaxies and \bl\ objects.  Using the observed X-ray luminosities of FR~I galaxies and \bl\ object is therefore not appropriate in the context of the jet model.   Plus, X-ray emission may be SSC or external inverse Compton (EC) for some \bl\ objects (see \S \ref{sec:kfcsdsshblsamp}).  The ``X-ray luminosities'' for the FR~I galaxies and \bl\ objects in the KFC sample are thus extrapolated from their optical nuclear luminosities, assuming an optically thin spectral index $\alpha_x$=0.6.  We explore the impact of this assumption in Sections~\ref{sec:mcsimFRIBL} and \ref{sec:bayessdss}.  Since synchrotron cooling is not important in the X-ray band for lower black hole masses, ``real'' observed X-ray luminosities are used for GBHs, Sgr~A$^{\star}$, and LLAGN.   
 
 \subsubsection{Doppler Boosting}
 \label{sec:kfcdoppler}
 Observed fluxes of  \bl\ objects are brightened by relativistic beaming.   Doppler boosting is negligible for other classes of AGN and GBHs considered here.  To first order, the level of Doppler boosting for \bl\ objects is similar in the radio and optical bands (we extrapolate \bl\ X-ray fluxes from the latter), so beaming does not significantly impact the inferred FP coefficients \citepalias{falcke04}. It may, however, contribute to some but not all of the intrinsic scatter about the plane \citep[e.g.,][]{li08}.   Since in this section we focus  on  statistics, we do not correct for Doppler beaming yet.  Such a discussion is deferred until \S \ref{sec:dr7blDoppler}.  We  refer the reader to \citet{heinz04_beaming} for a detailed discussion on how relativistic beaming can affect the FP, as well as \citet{giroletti06}. 
 
 \subsubsection{Error Budget and Intrinsic Scatter}
 \label{sec:errorbudget}
 We adopt the error budget explained in detail in \S 2.3 of \citetalias{koerding06}.  Black hole mass errors are typically of the order of $\sim$0.1~dex for GBHs, $\sim$0.35~dex for LLAGN and FR~I~galaxies (approximately the scatter about the $\mbh-\sigma_{\star}$ relation), and $\sim$0.46~dex for \bl\ objects (which have more indirect mass estimates).  Uncertainties on luminosities are dominated by errors in distance measurements, which typically range from 0.1--0.4~dex.   We also include $<$0.05~dex uncertainties on radio and X-ray flux measurements.   Additional uncertainties contribute to the intrinsic scatter about the plane, $\sigma_{int}$.  Potential sources of intrinsic scatter include the non-simultaneity of the AGN luminosity measurements,  relativistic beaming, not all sources having identical accretion rates, the environments surrounding each black hole differing from source to source, and the radio and X-ray wavebands not probing identical regions of different source's SEDs.  The last point means we do not observe exactly the same $\alpha_R$ and $\alpha_X$ for every source, and we therefore do not expect every source to follow exactly the same FP coefficients (see Figure~\ref{fig:coeff}).  Extrapolating ``X-ray luminosities'' from the optical for FR~Is and \bl s abates this effect some, but not entirely (see 
\S \ref{sec:bayessdss} for further discussion).

\subsection{Using the Merit Function to Regress the FP}
\label{sec:meritfcn}

Fitting the FP is a multivariate regression problem, where a variable, $y_i$, depends on two independent variables, $x_{ij}$, with $j=\{1,2\}$; each variable has a corresponding measurement error, $\sigma_{y_i}$ and $\sigma_{x_{ij}}$.  Previous authors used the Merit function,  a modified chi-square estimator \citep{press02}:
\begin{equation}
   \hat{\chi}^2 = \sum_i   %%
   \frac{ ( y_i - b - \sum_j a_j x_{ij} )^2} %   numerator
   {\sigma^2_{y_i} + \sum_j(a_j\sigma_{x_{ij}})^2}.   % denominator
\end{equation}

\noindent The unknown parameters, $a_j$ and $b$, are found by minimizing $\hat{\chi}^2$.  The constant $b$  can be solved for analytically (see Equation 3 of \citetalias{koerding06}), and the $a_j$ parameters can be found with a numerical optimization routine.

Here we associate $\log L_X$ with $y_i$, $\log L_R$ with $x_{i1}$, and $\log M$ with $x_{i2}$, and the unknown parameters $a_1$, $a_2$, and b  correspond to $\xi_R$, $\xi_M$, and $B$, respectively.  \citetalias{merloni03} assume measurement errors are isotropic (i.e., $\sigma_{L_X} = \sigma_{L_R} = \sigma_M$), and they adjust the uncertainties until $\hat{\chi}^2 = 1$.  This is an acceptable technique if the intrinsic scatter about the best-fit line dominates the error budget.  However, \citetalias{koerding06} show this may be a strong assumption, and incorrect values for $a_j$ and $b$ will be recovered if the errors are instead anisotropic.  \citetalias{koerding06} re-analyze the samples of \citetalias{merloni03} and \citetalias{falcke04}, properly accounting for errors in each measured variable.   They show the best-fit slopes of the FP strongly depend on the sample used, with the lowest-$\dot{m}$ AGN favoring the jet model.  Including AGN with higher accretion rates increases the intrinsic scatter, and then the inferred slopes are perhaps  more consistent with X-ray emission from inverse Compton.  However, uncertainties on the best-fit coefficients from the Merit function regression in \citetalias{koerding06} do not unambiguously identify the proper emission mechanism.

The Merit function has some limitations.  For example, it cannot handle coupled uncertainties, which is a potentially severe limitation since radio and X-ray luminosities are both dominated by errors in  distance measurements.  Also,  while \citetalias{koerding06} properly account for measurement errors in each variable, their estimation of $\sigma_{int}$  is inferred by adjusting $\sigma_{int}$ until $\hat{\chi}^2$ is unity.  We thus re-regress the FP with a more sophisticated Bayesian technique (see next section) that can handle the above limitations (as well as large measurement uncertainties) to see if the results of \citetalias{koerding06} can be improved.

\subsection{Bayesian Regression}
\label{sec:kelly07}
We use the method of \citet[][hereafter \citetalias{kelly07}]{kelly07} to regress the FP.   \citetalias{kelly07} take a Bayesian approach toward linear regression, estimating the probability distribution of the FP parameters given the measured data.\footnote{{\tt mlinmix\_err.pro} in the IDL Astronomy User's Library -- \url{http://idlastro.gsfc.nasa.gov/}. } %%
  The method assumes that the measurement errors are normal, that the intrinsic scatter about the FP is normal, and that the distribution of the independent variables can be approximated as a mixture of Gaussian functions. Under these assumptions it then uses a Markov chain Monte Carlo (MCMC) method to simulate random draws of the FP parameters from their probability distribution, given the measured data (i.e., the `posterior' probability distribution). These random draws for the FP parameters then allow us to estimate their best-fit values,  standard errors, and probability distributions. We refer the reader to \citetalias{kelly07}, especially their Sections~4 and 6, for more details.

   In this work, we typically take 10$^4$ random draws from the posterior.  Unless stated otherwise, we quote ``best-fit'' values of $\xi_R$, $\xi_M$, $B$, and $\sigma_{int}$ as the median value of each quantity's posterior distribution.  After checking that each posterior distribution follows an approximately normal distribution, we report uncertainties  in the text as the $\pm$1$\sigma$ standard deviations . Since the best-fit regression coefficients are correlated, we will draw error ellipses in figures at both the 1$\sigma$ (68\%) and 3$\sigma$ (99.7\%) levels.
     
   In particular, this Bayesian method overcomes the two Merit function limitations described in the previous section. For one, the method of \citetalias{kelly07} can  account for correlated errors.    Secondly, it samples posterior distributions not just for the unknown parameters $\xi_R$, $\xi_M$, and $B$, but also for the intrinsic scatter $\sigma_{int}^2$.  That is, $\sigma_{int}$ is estimated directly from the data, and not by adjusting its value by hand until a modified chi-square estimator is close to unity.  Especially for a correlation with relatively large intrinsic scatter like the FP, this has the attractive quality of not overweighting data points with small measurement errors that are highly scattered from the best-fit line.\footnote{Note, as  for the Merit function, if measurement errors are over- or under-estimated, then $\sigma_{int}^2$ is still under- or over-estimated, respectively.}  %%
    The method of \citetalias{kelly07} is also very well suited for handling heterogeneously selected datasets with large measurement uncertainties, like the KFC sample, as long as the intrinsic distributions of the independent variables can be modeled as a mixture of Gaussians.  This is a reasonable assumption as long as the number of Gaussians is large (the Gaussians do not need to be physically meaningful).  Also, the measurement errors in the independent variables do not need to be of similar magnitude.

 \subsection{Reliability of the Bayesian Regression}
 \label{sec:mcsim}
\begin{figure}
  \includegraphics[scale=0.85]{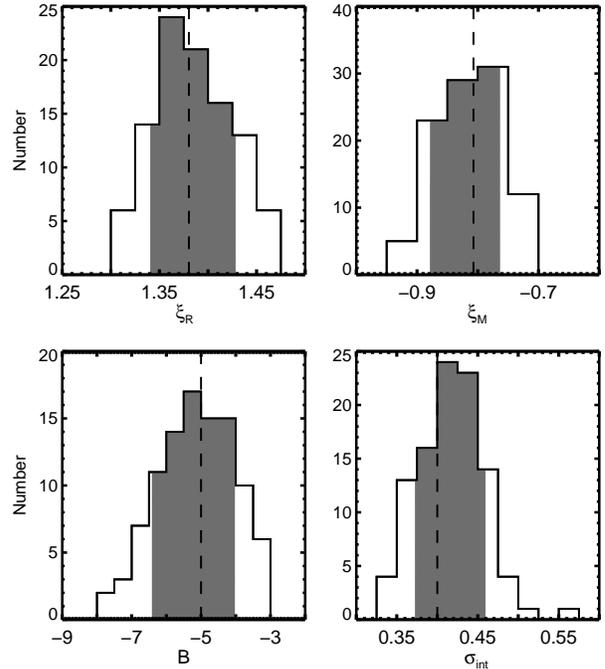}
 \caption{Distributions of best-fit parameters with the Bayesian regression on 100 simulated FPs (\S\ref{sec:mcsim}). The input values ($\xi_R$, $\xi_M$, $B$, $\sigma_{int}$)=(1.38,-0.81,-5.00,0.40) are shown with dotted lines.  The shaded regions illustrate the 16-84\% confidence intervals (1$\sigma$). }
 \label{fig:mcsim}
 \end{figure}

 Before directly comparing  Bayesian regressions using the technique of \citetalias{kelly07} to the Merit function regressions in \citetalias{koerding06}, we test the accuracy of the Bayesian technique on simulated FPs (also see \citetalias{kelly07} for more details on its reliability and comparison to other regression techniques).   We take the observed radio luminosities and black hole masses for each source in the full KFC sample (77 data points),  assume these are the ``true" values for each accreting black hole, and we predict ``true'' X-ray luminosities assuming ($\xi_R$, $\xi_M$, $B$) = (1.38, -0.81, -5.00) and then add $\pm$0.4~dex intrinsic scatter.   The input FP slopes correspond to $\alpha_R=-0.15$ and $\alpha_X=0.6$ (i.e., p=2.2) in the jet model (see Figure~\ref{fig:coeff}).   Fake noise is then randomly added to each ``true'' value of $L_R$, $\mbh$, and $L_X$ by assuming their errors are Gaussian with a standard deviation equal to the observed errors.   When adding simulated noise, we take into account the fact that uncertainties in $L_R$ and $L_X$ are correlated.     We then regress the simulated FP with the \citetalias{kelly07} Bayesian method, taking the best-fit parameters as the median values of the posterior distributions for $\xi_R$, $\xi_M$, $B$, and $\sigma_{int}$.    We repeat this exercise 100 times (i.e., we regress 100 simulated FPs), and the distributions of the best-fit parameters are shown in Figure~\ref{fig:mcsim}.
 
The average best-fit parameters of the 100 simulated FPs match very well to the ``true'' input parameters, we recover $\left<\xi_R\right> = 1.38\pm 0.04$, $\left<\xi_M\right>=-0.81 \pm 0.05$, $\left< B\right>=-5.12 \pm 1.10 $, and $\left<\sigma_{int}\right>=0.42 \pm 0.04$~dex (median values of the fitted parameters are similar to the above means).   The uncertainties quoted above are the standard deviations of the distributions shown in Figure~\ref{fig:mcsim}.  These uncertainties are consistent with the average of the 1$\sigma$ errors measured during the regression of each simulated FP: $\left<\sigma_{\xi_R}\right> = \pm 0.04$, $\left<\sigma_{\xi_M}\right>=\pm 0.06$, $\left< \sigma_B\right>= \pm 1.16 $, and $\left<\sigma_{\sigma_{int}}\right>= \pm 0.04$.  The Bayesian technique thus recovers accurate coefficients and intrinsic scatter about the FP, as well as realistic uncertainties in each parameter provided our adopted error budget is reasonable.  

\subsubsection{Synchrotron Cooling and Extrapolating FR~I and \bl\ X-ray Luminosities From the Optical}
\label{sec:mcsimFRIBL}
\begin{figure}
  \includegraphics[scale=0.88]{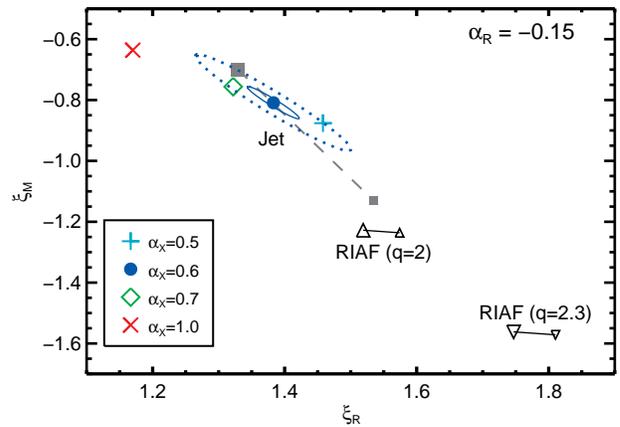}
 \caption{The median value of the 100 best-fit coefficients in the Monte Carlo simulations assuming $\alpha_X$=0.5, 0.6, 0.7, and 1.0 (cyan plus sign, blue filled circle, green diamond, and red cross, respectively) to extrapolate FR~I and \bl\ X-ray luminosities from observed optical nuclear luminosities.  1$\sigma$ (solid blue line) and 3$\sigma$ (dotted blue line) error ellipses are shown for  $\alpha_X=0.6$; errors are similar for the other three cases, but their error ellipses are not shown for clarity.  Overplotted for reference are the theoretical predictions for $\xi_M$ and $\xi_R$ for the optically thin jet synchrotron and RIAF models in the case $\alpha_R = -0.15$. The symbols showing the theoretical jet and RIAF predictions (i.e., squares and triangles connected by lines) have the same meaning as in Figure~\ref{fig:coeff}). The simulated FPs assume a jet model with  $\alpha_R=-0.15$ and $p=2.2$.  Extrapolating X-ray luminosities from the optical can  impact the best-fit coefficients.  However, the effect (when $\alpha_r = -0.15$) is not severe enough to infer the incorrect radiation mechanism for X-rays from the best-fit coefficients, especially for very inefficient (i.e., large $q$) RIAFs.  This figure appears in color in the online version of the article.}  
 \label{fig:mcsimDiffp}
 \end{figure}

Extrapolating X-ray luminosities from optical nuclear luminosities for FR~I galaxies and \bl\ objects could systematically bias the regressions.  For example, the KFC sample assumes $\alpha_X =0.6$, but using $\alpha_X=0.5$ or $\alpha_R=0.7$ would increase or decrease  the FR~I and \bl\ ``X-ray luminosities'' by about a factor of two, respectively.    To test the importance of this effect, we take the 100 above simulated planes (which were simulated assuming $\alpha_R=-0.15$ and $\alpha_X=0.6$ given the jet model).     We then increase the FR~I and \bl\ X-ray luminosities by a factor of two to simulate extrapolating their luminosities with $\alpha_X = 0.5$ and re-regress each simulated plane.  We then repeat but decreasing their X-ray luminosities by a factor of two to simulate an extrapolation assuming $\alpha_X=0.7$.     The effect  on the coefficients $\xi_R$ and $\xi_M$ from extrapolating with different values of $\alpha_X$ is shown in Figure~\ref{fig:mcsimDiffp}.  Although the median best-fit coefficients of the 100 simulations are formally consistent within $\pm$3$\sigma$, larger values of $\alpha_X$ appears to bias the FP toward shallower slopes.  This is expected since steeper (i.e., larger $\alpha_X$) power laws will predict lower X-ray luminosities, so the correlation looks flatter.  Similar intrinsic scatters, $\sigma_{int}=0.42$--0.43~dex (with associated uncertainties of $\pm$0.04~dex), are measured for all three values of $\alpha_X$.
  
  Finally,   we simulate the situation where X-ray telescopes observe optically thin synchrotron jet emission with $p=2.2$ (still assuming the radio is optically thick with $\alpha_R=-0.15$) for lower mass ($<$10$^8$~$\msun$) black holes, but  X-rays from higher mass black holes are synchrotron cooled and follow $\alpha_X=1$.   We test this by lowering the FR~I and \bl\ X-ray luminosities by a factor of 12 and re-run 100 simulations.  From Figure~\ref{fig:mcsimDiffp}, we see this pushes the coefficients to even shallower slopes.  We may have also measured an increase in intrinsic scatter, $\sigma_{int}=0.50\pm0.04$~dex, which is not surprising.

In \S \ref{sec:fitBL} we will confirm the assertions of \citetalias{falcke04} and \citetalias{koerding06} that extrapolating X-ray luminosities from lower frequencies is necessary if X-rays are dominated by optically thin jet synchrotron (i.e., one cannot test the theoretical FP coefficients with observed X-ray fluxes for the most-massive SMBHs because synchrotron cooling is too strong).  However, our simulations show one must be careful in how this extrapolation is made.  In the case of the KFC sample, if extrapolating with $\alpha_X=0.6$, then the optical flux should be measured at a frequency where the non-thermal emission actually follows a local power law spectrum $f_{\nu}\sim \nu^{-0.6}$, and therefore SED modeling is necessary (see \S \ref{sec:fitBL}).  We note, however, while this extrapolation effect is not negligible, in certain cases it may not be severe enough to strongly influence the inferred X-ray radiation mechanism.  For example, the KFC sample includes primarily low-accretion rate black holes with flat radio spectra $\alpha_R\sim-0.15$.    So, if regression of the KFC sample were to yield coefficients (and associated uncertainties) similar to our simulations, then it would likely still be possible to exclude the RIAF models in all four cases (especially the ones with very low-efficiency, i.e., $q\sim2.3$).  Regardless, this is a potential  source of systematic bias one should be concerned about, demanding special attention to how one analyzes multiwavelength data before inclusion in FP studies.

\subsection{Bayesian Regression of the KFC Sample}
\label{sec:kfcbayes}
\begin{figure*}
\includegraphics[scale=1.0]{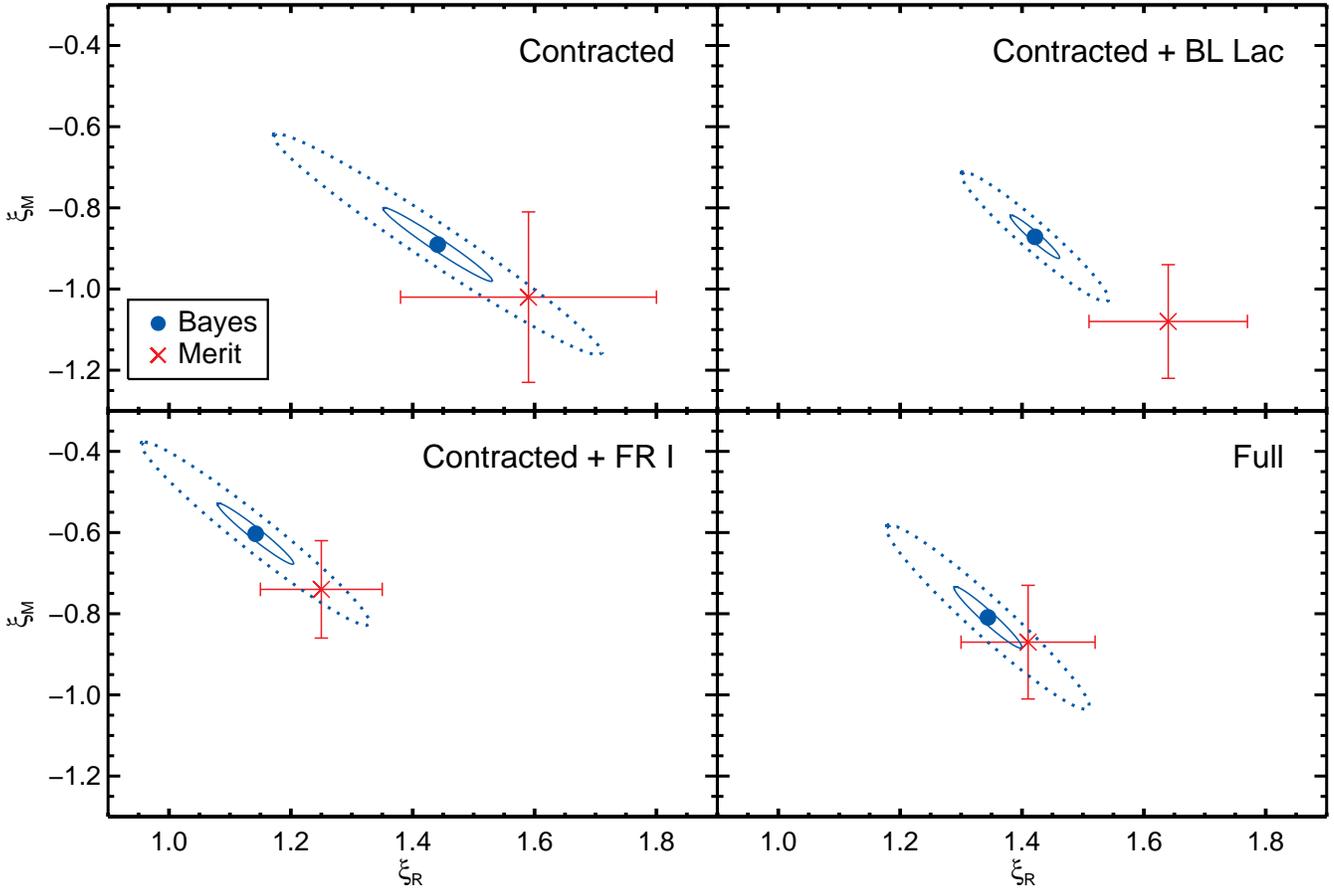}
\caption{Results of the Bayesian regressions (blue circles)  compared to Merit function regressions (red crosses) from \citetalias{koerding06} for the four KFC subsamples considered here.   1$\sigma$ (solid blue line) and 3$\sigma$ (dotted blue line) error ellipses are drawn for the Bayesian regression, and 1$\sigma$ error bars are shown for the Merit function.  Note, the 1$\sigma$ Bayesian uncertainties are  smaller than the 1$\sigma$ Merit function uncertainties.    The Merit function seems to systematically recover steeper slopes than the Bayesian regression.   Synchrotron cooling moves the contracted+FR~I best-fit slopes toward shallower values.   This figure appears in color in the online version of this article. }
\label{fig:meritvbayes}
\end{figure*}

Here, we apply the Bayesian regression on the contracted, contracted+\bl, contracted+FR~I, and full KFC samples (see \S\ref{sec:kfc} and Table~\ref{tab:samples} for sample definitions), and we compare to the regressions in \citetalias{koerding06}  using the Merit function.  Results are summarized in Table~\ref{tab:kfcmeritVbayes} and  Figure~\ref{fig:meritvbayes}.    Save for the contracted+\bl\ sample, the Merit function regressions are consistent with the Bayesian regression at the $3\sigma$ level. However, there are important differences between the results from the two methods.  First,  the slopes recovered by the Merit function are consistently steeper than those from the \citetalias{kelly07} method.    We attribute some of this  to the Bayesian regression properly handling correlated measurement errors between radio and X-ray luminosities.    Also, the Bayesian regression  recovers larger best-fit intrinsic scatters than the Merit function (note, the Merit function does not actually measure $\sigma_{int}$ directly from the data, rather its value is adjusted until $\hat{\chi}^2 \sim 1$).    Finally, uncertainties on the best-fit regression coefficients are consistently smaller with the Bayesian approach, which may eventually allow more definitive statements on the most-likely X-ray emission mechanism.  We note, however, quoted uncertainties on the correlation coefficients do not account for the potential biases from extrapolating FR~I and \bl\ X-ray luminosities from the optical discussed in \S \ref{sec:mcsimFRIBL} and Figure~\ref{fig:mcsimDiffp}.

\begin{table*}
\caption{Bayesian vs.\ Merit Function Regression of KFC Sample}
\label{tab:kfcmeritVbayes}
\scriptsize
\begin{tabular}{l c cc r@{.}l c cc r@{.}l c }
\hline
		                     & %1
		                     & %2
		 \multicolumn{5}{c}{Bayesian Regression Coefficients (this work)}   & %3-6 
		 \multicolumn{5}{c}{Merit Function Coefficients (\citetalias{koerding06})}  \\         
		   \cline{3-7}       \cline{8-12}
		Sample      &  %1
	         Number           &  %2           
	         	$\xi_R$  	   &   %3
		$\xi_M$     &  %4
		\multicolumn{2}{c}{$B$}             & %5
		$\sigma_{int}$ & %6
		$\xi_R$      & %7
		$\xi_M$      &	%8
		\multicolumn{2}{c}{$B$}            & %9
		$\sigma_{int}$ \\ %10

\hline
       Contracted &                43 &   1.44 $\pm$ 0.09 &  $-$0.89 $\pm$ 0.09 &  $-$5&95 $\pm$ 2.58 &   0.40 $\pm$ 0.06 &   1.59 $\pm$ 0.21 &  $-$1.02 $\pm$ 0.21 & $-$10&15 $\pm$ 6.17 &             0.12 \\ 
Contracted+BL Lac &                62 &   1.42 $\pm$ 0.04 &  $-$0.87 $\pm$ 0.05 &  $-$5&37 $\pm$ 1.15 &   0.34 $\pm$ 0.05 &   1.64 $\pm$ 0.13 &  $-$1.08 $\pm$ 0.14 & $-$11&67 $\pm$ 3.59 &             0.18 \\ 
  Contracted+FR I &                58 &   1.14 $\pm$ 0.06 &  $-$0.60 $\pm$ 0.08 &   2&65 $\pm$ 1.78 &   0.48 $\pm$ 0.05 &   1.25 $\pm$ 0.10 &  $-$0.74 $\pm$ 0.12 &  $-$0&46 $\pm$ 2.93 &             0.28 \\ 
             Full &                77 &   1.34 $\pm$ 0.06 &  $-$0.81 $\pm$ 0.08 &  $-$3&13 $\pm$ 1.58 &   0.58 $\pm$ 0.05 &   1.41 $\pm$ 0.11 &  $-$0.87 $\pm$ 0.14 &  $-$5&01 $\pm$ 3.20 &             0.38 \\ 
\hline
\end{tabular}
\end{table*}

The best-fit regression coefficients from the Bayesian method for the contracted, contracted+\bl\ and full KFC samples seem less sensitive to the subsample than for the Merit function.  The regression coefficients for the contracted+FR~I subsample, however, are much shallower.   The subsamples including FR~I galaxies also have the largest intrinsic scatter about the plane.  We attribute this to several systematics related to difficulties in making the requisite observations of FR~I galaxies for FP studies (see \S \ref{sec:whyfribad}).    That is, FR~I galaxies bias the best-fit slopes (but the inclusion of \bl\ objects in the full sample seems to partially offset this bias).  FR~I galaxies should be removed from the KFC sample until these systematics are more rigorously addressed (although see \citealt{hardcastle09} regarding the placement of unbeamed radio galaxies onto the FP).

\subsubsection{Systematic Challenges to Including FR~I Galaxies}
\label{sec:whyfribad}
 
From a purely theoretical standpoint, FR~I galaxies should follow the  FP like other low-accretion rate black holes.    They have sub-Eddington accretion rates, and, especially as $\dot{m}$ decreases,  their broad-band SEDs (including X-ray emission) are generally jet-dominated \citep[e.g.,][]{evans06}.  However, there are several observationally based reasons to suspect that FR~I galaxies could bias the FP regression.  Most important is the effect of synchrotron cooling.   For \bl\ objects, Doppler beaming will push the synchrotron cutoff  toward higher observed frequencies, so that the optical waveband still probes optically thin jet emission for many \bl\ objects.  However, for unbeamed sources like FR~I galaxies, even the optical could already be synchrotron cooled (also see \S \ref{sec:bayessdss}).  Thus, the $\alpha_X=0.6$ assumption for extrapolating FR~I ``X-ray luminosities'' from the optical will introduce a larger bias than for \bl\ objects.  Synchrotron cooling will thus systematically move the best-fit coefficients toward shallower values.  Interestingly, the $\xi_R$ and $\xi_M$ values of the contracted+FR~I subsample are similar to our Monte Carlo simulations where FR~I and BL~Lac ``X-ray luminosities'' are estimated using $\alpha_X=1$ (to simulate synchrotron cooled jet emission in the optical, Figure~\ref{fig:mcsimDiffp}).

Another potential source of systematic bias introduced by  FR~I galaxies is that it can be harder to isolate their nuclear radio emission, since their radio cores do not appear as bright as \bl\ objects.  The majority of FR~I galaxies in the KFC sample have only VLA observations \citep{chiaberge99}, meaning there is potentially some concern for contamination to their nuclear luminosities from optically thin radio emission that appears point-like at VLA resolution.  (Of lesser, but perhaps not negligible, concern is radio emission from star formation contributing to the observed nuclear radio flux).  Thus, FR~I radio spectra could be steeper than the rest of the objects in the KFC sample, which will also bias the best-fit regressions toward shallower slopes.  Such systematics are likely less important than their synchrotron cutoff occurring below the optical waveband, especially since most of the KFC FR~I galaxies are at low-redshift ($z<0.1$).  We mention such potential systematics here primarily to illustrate that, in general, \bl\ objects provide ``cleaner'' probes of jet emission.  

Since \bl\ objects and FR~I galaxies are presumed to be the same type of object  but at different orientations, we prefer using \bl\ objects over FR~I galaxies in this work.  To properly include FR~I galaxies, one would need to ensure observed radio core fluxes are only including synchrotron jet emission, and  SED modeling would be necessary to verify ``X-ray luminosity'' estimates are not influenced by dynamic cooling of the synchrotron emitting particles.  

\subsection{Jet Synchrotron vs.\ Inverse Compton X-ray Emission}
\label{sec:jetvscorona}
The best-fit slopes of the FP can be used to investigate if X-ray emission from sub-Eddington black holes is dominated by optically thin jet synchrotron or inverse Compton. However, we implicitly assume the jet interpretation when extrapolating FR~I and \bl\ X-ray luminosities from the optical (such an extrapolation is improper if inverse Compton).  So we must omit  the most massive black holes from this discussion, and we therefore only consider  the contracted KFC subsample here.  Even though the contracted subsample contains the smallest number of objects, and it has the most limited dynamic range in black hole mass and luminosity, the uncertainties on the Bayesian regression coefficients are narrow enough to make definitive statements on the dominant X-ray emission mechanism.  We stress again that, in reality, both jet synchrotron and coronal inverse Compton components are likely present in all sub-Eddington black holes.  The following discussion is thus aimed toward determining which emission mechanism, on average, dominates the  X-rays observed from low-accretion rate black holes.

\begin{figure*}
\includegraphics[scale=1.0]{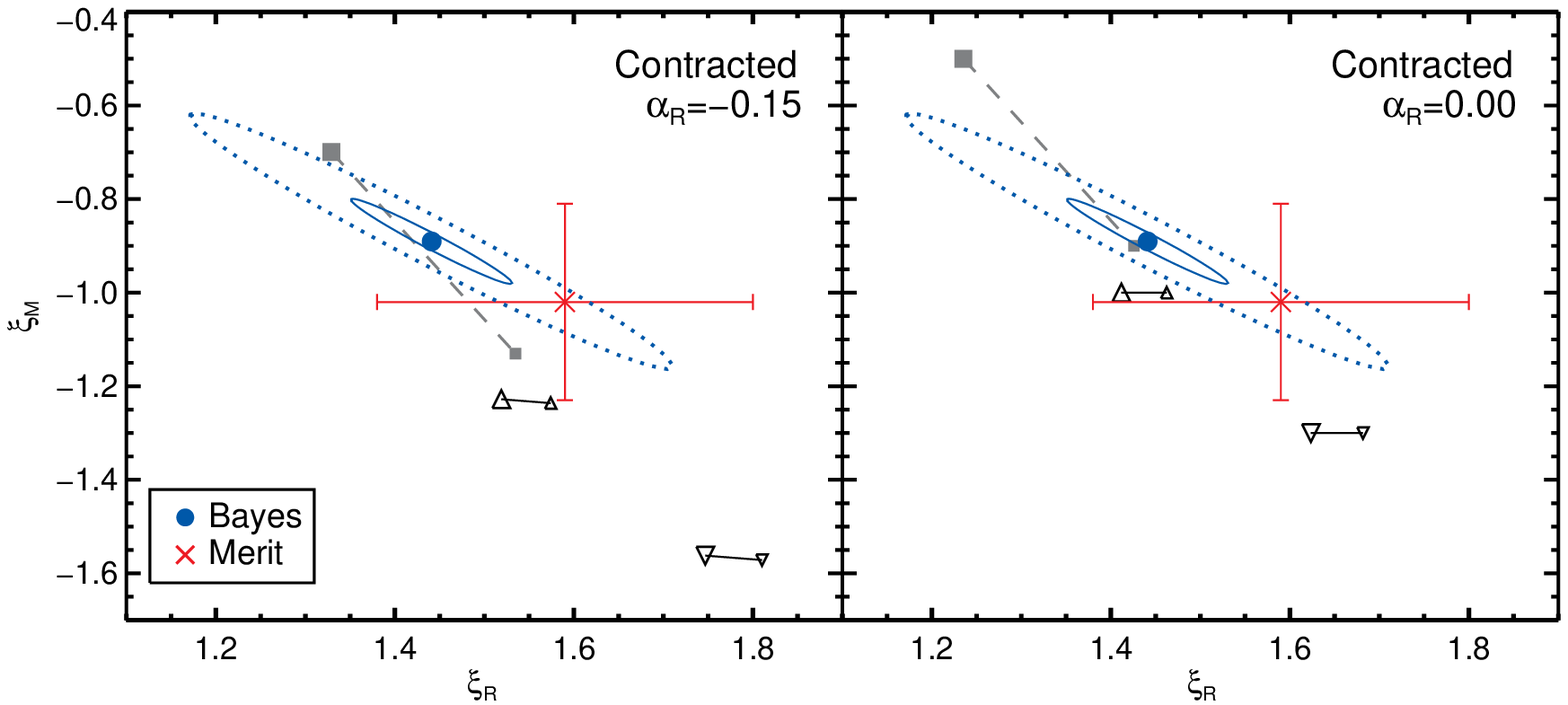}
\caption{Bayesian and Merit function regression coefficients for the contracted KFC subsample, compared to theoretical predictions for the jet (squares) and RIAF/corona models (triangles for q=2.0; upside down triangles for q=2.3) from Equations~\ref{eq:jetcoeff} and \ref{eq:coronacoeff}.     We separate predicted FP coefficients for $\alpha_R=-0.15$ and 0.00 into different panels for clarity.  The Bayesian regression of the contracted subsample favors the jet synchrotron model. This figure appears in color in the online version of this article. }
\label{fig:meritVbayesContracted}
\end{figure*}

Most (and perhaps all) accreting  black holes  in the KFC sample have $\alpha_R \le 0$, so we compare the contracted subsample regression coefficients to the predicted coefficients only for flat or inverted radio spectra.  In Figure~\ref{fig:meritVbayesContracted}, we compare the best-fit regression coefficients to the  coefficients that are predicted for $\alpha_R=-0.15$ and 0.00, in the cases where X-rays are dominated by jet synchrotron, or by inverse Compton off a corona with radiative efficiencies q=2.0 or 2.3.  Considering even just 1$\sigma$ error bars on the Merit function regression coefficients, one cannot conclusively exclude any of the three models.  However, the Bayesian regression favors the coefficient predictions if X-rays are dominated by optically thin jet synchrotron.   As long as $\alpha_R \leq 0.0$, the q=2.3 RIAF/corona model is excluded at the $>$3$\sigma$ (i.e., $>$99.7\%) level.  The Bayesian regression excludes the q=2.0 RIAF/corona model with similarly high-confidence if the majority of objects in the KFC have inverted radio spectra (which is likely the case), but perhaps less significantly ($p\sim0.997$) if the majority of objects have perfectly flat radio spectra (i.e., the Bayesian regression's 3$\sigma$ error ellipse is similar to the expected RIAF coefficients if $q=2.0$, $\alpha_R=0.00$, and $p\sim2.8-3.0$).  We thus conclude that the average black hole in the contracted KFC sample emits most of its X-rays as optically thin synchrotron radiation, and we exclude very radiatively inefficient RIAFs (i.e., q=2.3) with high-confidence.

None of the theoretical predictions assuming $\alpha_R=0.5$ is consistent with the best-fit regressions, as expected since all KFC objects have flatter radio spectra.  Also as expected, radiatively efficient accretion flows are excluded.  However, if we compare to predicted coefficient values assuming $\alpha_R=0.15$, then the Bayesian regression analysis no  longer excludes any of the three models.   While the majority of KFC black holes indeed have $\alpha_R<0$, a small number of LLAGN may have radio spectral indices extending up to $\alpha_R=0.3$ based on non-simultaneous multifrequency radio imaging (see \citealt{terashima03} and \citealt{nagar05}, from which the KFC LLAGN are taken).    So, we searched NED\footnote{The NASA/IPAC Extragalactic Database (NED) is operated by the Jet Propulsion Laboratory, California Institute of Technology, under contract with the National Aeronautics and Space Administration.} %%
for all publicly available radio photometric data for the 17 KFC LLAGN.  We find only 4/17 KFC LLAGN could potentially have $\alpha_R>0$, meaning $>$90\% of the 43 objects in the  contracted KFC subsample have $\alpha_R<0$.   Thus, the presence of a small number of LLAGN potentially with $\alpha_R>0$ may contribute to the intrinsic scatter about the FP,  but they  unlikely represent a large enough population to severely influence the best-fit (i.e., average) regression coefficients.   We conclude that it is a reasonable assumption that our sample has flat/inverted radio spectra.  However,  it would be worth confirming our conclusion in future work with simultaneous multifrequency radio imaging.

It is important to once again stress that our conclusion favoring the synchrotron X-ray  model applies specifically to the case of low-accretion rate black holes with flat/inverted radio jets.  It is not appropriate to extrapolate this result to higher accretion rate \smbh s, which likely have X-rays dominated by emission from a corona.   A similar study focusing only on black holes at higher accretion rates is  out of the scope of this paper.

\section{Fitting Very Massive Black Holes onto the Plane}
\label{sec:fitBL}
Our re-analysis of the KFC sample shows that one must overcome more systematics to obtain ``X-ray" measurements from the most massive accreting SMBHs ($>\approx 10^8 \msun$).  In this section,  we shift our focus toward better understanding and quantifying these systematics.   As described below, the inclusion of  very massive SMBHs when regressing the FP will not yield additional insight into X-ray radiative processes from the lower mass black holes.  However, many previous FP studies include AGN with very massive central black holes, so  an investigation focusing on potential biases introduced by those AGN is extremely important.

We begin by removing the \bl\ objects from the KFC sample and we replace them with a uniform sample of \bl\ objects from the Sloan Digital Sky Survey \citep[SDSS;][]{york00} with  black hole mass estimates \citep{plotkin11}.  We then re-regress the FP with the same Bayesian technique as in the previous section. We exclude FR~I galaxies in this section for the reasons described in \S\ref{sec:whyfribad} (also see \S \ref{sec:whyusebl}). The  Bayesian regression technique in conjunction with our recent and \textit{uniform} SDSS \bl\ sample affords the requisite sensitivity to finally perform  a rigorous study on biases introduced by the largest black hole mass bin.

\subsection{Why Use \bl\ Objects to Represent the High-end of the Mass Spectrum?}
\label{sec:whyusebl}
First, it is important to reiterate the philosophical challenge to including very massive black holes in FP studies.  In the previous section, we excluded the most massive KFC black holes (i.e., FR~I galaxies and \bl\ objects) from our discussion when attempting to determine the dominant X-ray radiation mechanism.  The reason is that, for lower mass black holes, one can use real X-ray luminosities  to differentiate between the optically thin synchrotron and inverse Compton scenarios.   However, for the most massive black holes, synchrotron emission in the X-ray waveband is generally radiatively cooled (or sometimes SSC/EC).   So, even ``jet'' dominated SEDs from very massive black holes will not follow the theoretical predictions of Equation~\ref{eq:jetcoeff}, if one uses ``real X-ray" data.   However, if very massive black hole SEDs are \textit{not} jet dominated, then X-rays would still be sensitive to inverse Compton emission and real X-ray data should be used to place them on the FP.   Thus,  one must make an assumption on the dominant radiative process from the most massive black holes \textit{before} deciding from which waveband to estimate their ``X-ray luminosities."   This \textit{a priori} assumption on the radiative mechanism then makes it impossible to use the most massive black holes to diagnose the dominant radiation mechanism from a statistical black hole sample spanning the entire mass spectrum.\footnote{Although theoretical FP coefficients for synchrotron cooled X-rays are predicted by \citet{heinz04}, comparing observations to those predictions presents a similar problem: observations in different wavebands are necessary to uniformly probe synchrotron cooled emission across the entire mass scale.}   %%  

Even with the restricted  dynamic range of the contracted KFC subsample, the Bayesian regression technique is sophisticated enough to infer that  X-rays from those 43 objects are generally dominated by optically thin synchrotron emission.  To study in detail the  systematics introduced by including the most massive black holes, we must add AGN with $>\approx 10^8~\msun$ central black holes back into the sample; we should only consider AGN that are very massive analogs to the types of accreting black holes in the contracted KFC subsample.    \bl\ objects are the best very massive  analogs: \bl\ objects have sub-Eddington accretion rates \citep[e.g.,][]{ghisellini09} and  (uncontroversially) jet-dominated SEDs with flat/inverted radio spectra.   In principle we could also consider FR~I galaxies, but we choose to restrict our massive SMBHs exclusively to \bl\ objects.  The primary reason is the Doppler boosting of \bl\ jets: we can better isolate their jet core radio emission (even without VLBI observations), and synchrotron cooling kicks in at slightly higher observed frequencies.  The latter makes  it easier to estimate optically thin synchrotron jet  luminosities for \bl\ objects compared to FR~I galaxies.  However, we refer the reader to \citet{hardcastle09} for a FP study that overcomes some of the above systematics: with X-ray spectral modeling, \citet{hardcastle09} are able to compare accretion-related vs. jet-related X-ray emission for a sample of unbeamed radio galaxies.

In summary, by better controlling systematics at the high-mass end with a uniform sample of \bl\ objects, we can  investigate how very  massive black holes may  bias  FP regressions.  Since the most massive SMBHs provide the widest dynamic, we will in the process also derive the most accurate FP coefficients to date \textit{in the context of the jet model.}  This assumption of the jet model is extremely important, as our handling of multiwavelength data for \bl\ objects is applicable only if their SEDs are jet dominated.  However, this assumption is also very reasonable in our case.  We showed in the last section that the lower mass contracted KFC black holes are indeed dominated by  synchrotron emission, and it is very well-established that \bl\ SEDs  are jet dominated.  That is, we know what \bl\ broadband SEDs ought to look like, allowing us to better investigate how the most massive SMBHs can  bias FP regressions.

\subsection{Why Replace the KFC \bl\ Objects with Ones From the SDSS?}
\label{sec:whyreplacekfcbl}
Instead of using the \bl\ objects from the KFC sample, we opt to replace them with \bl\ objects from the SDSS.  While \citetalias{koerding06} do  keep the KFC sample as uniform as possible, it is by nature heterogeneous, and this is especially true for their \bl\ objects.    The 19 \bl\ objects in the KFC sample were taken from \citet{sambruna96}, and they were originally discovered either by the X-ray selected sample from  the {\it Einstein} Observatory Extended Medium-Sensitivity Survey \citep[EMSS;][]{morris91, stocke91} or by the 1 Jy radio-selected sample \citep{stickel91}.   These venerable \bl\ samples composed the first complete sets of \bl\ objects, and they provided tremendous advances toward our understanding of \bl\ phenomena.  However, their shallow flux limits (and in the case of the EMSS, relatively small areal coverage) produced biased samples that are not representative of the actual \bl\ population \citep[e.g.,][]{laurent98, plotkin08}.   Furthermore, only \bl\ objects for which \citetalias{falcke04} found black hole mass measurements in the literature made it into the KFC sample (and these black hole masses were also derived non-uniformly).

Since \bl\ objects provide such a large lever arm for regressing the FP, it is important to reduce as many of the above biases as possible to minimize the observational systematics discussed in \S \ref{sec:whyusebl}.  That is why we replace the KFC \bl\ objects with ones from the SDSS.  Although \bl\ catalogs from the SDSS are not free of selection biases, these samples are large enough to include a population of objects more representative of the parent population \citep[][hereafter \citetalias{plotkin10}]{collinge05,plotkin10}.  Perhaps more importantly in our opinion, the sample is uniform.  That is,  all objects are selected the same way, the multiwavelength data are taken with the same telescopes, and  black hole measurements are derived with the same technique.

\subsection{The SDSS BL~Lac Objects}
\subsubsection{The KFC+SDSS Sample}
\label{sec:kfcsdsssamp}
We use \bl\ objects from the large (723~object) optically selected catalog from the SDSS \citepalias{plotkin10}.  \citetalias{plotkin10} select \bl\ candidates based on their featureless optical spectra, applying the standard criteria of \ion{Ca}{ii}~H/K breaks smaller than 40\% and no emission lines with rest-frame equivalent widths stronger then 5~\AA\ \citep[see, e.g.,][]{stocke91,landt02}.  For this FP  study, we require radio and X-ray luminosities, as well as central black hole mass measurements; we select a subset of \nblmass\ \bl\ objects for which we have data for all three FP axes.   We refer to this sample as the \textit{KFC+SDSS sample}  (KFC GBHs, Sgr~A$^\star$, LLAGN, and SDSS \bl s; 98~objects; see Table~\ref{tab:samples}).

\subsubsection{The KFC+SDSS-HBL and KFC+SDSS-LBL Samples}
\label{sec:kfcsdsshblsamp}
It is well known, e.g., from the blazar sequence \citep{fossati98}, that \bl\ object SEDs can vary drastically from object to object, with their cutoff frequencies occurring anywhere from the near-infrared to the soft X-ray \citep[e.g.,][]{nieppola06}.  \bl\ objects with synchrotron cutoff frequencies in the near-infrared are  referred to as low-energy cutoff \bl\ objects (LBLs), those with soft X-ray cutoff frequencies  are called high-energy cutoff \bl\ objects (HBLs), and intermediate-energy cutoff \bl\ objects (IBLs) are in between \cite[e.g.,][]{padovani95_apj}.    In addition to synchrotron emission, \bl\ objects also emit at higher frequencies via SSC and possibly EC processes.   LBL/IBL X-ray emission is typically already SSC/EC \citep[e.g.,][]{abdo10_sed}, and it may be difficult (with extant data) for us to reliably estimate the luminosity of optically thin synchrotron emission for LBLs.    However, this is much easier to do for HBLs, where X-ray emission is still synchrotron, albeit radiatively cooled.

With such a large parent sample of \bl\ objects, we can consider only HBLs and still retain a sufficiently large number of \bl\ objects.    We thus also create the KFC+SDSS-HBL subsample, which consists of the 43 contracted KFC black holes and 39/55 SDSS \bl\ objects that we classify as HBLs (82 objects total; see Table~\ref{tab:samples}).   We similarly create the KFC+SDSS-LBL sample, which only adds the 16 IBL/LBLs to the contracted KFC subsample (59 objects total; see Table~\ref{tab:samples}).  We define HBLs  based on the ratio of  X-ray to radio emission \citep[e.g.,][]{padovani95_apj, perlman96}.  \citetalias{plotkin10} followed relatively standard convention, and they classified \bl\ objects with $\alpha_{rx} < 0.75$ as HBLs.\footnote{$\alpha_{rx}$ is the broadband X-ray to radio spectral index: $\alpha_{rx}=-\log\left (L_{\nu,1~keV}/L_{\nu,5~GHz}\right)/7.68$, where $L_{\nu,1~keV}$ and $L_{\nu, 5~GHz}$ are X-ray and radio specific luminosities at rest-frames 1~keV and 5~GHz, respectively.}  %%
  Here, we adopt a more conservative cut, and we identify our 39 HBLs as objects with $\alpha_{rx}<0.7$ (we calculate $\alpha_{rx}$ values from the rest-frame 1~keV X-ray and 5~GHz radio luminosity densities in Table~7 of \citetalias{plotkin10}).  The remaining 16 \bl\ objects with $\alpha_{rx}>0.7$ are then IBLs and LBLs.    The reason for our more stringent limit on X-ray brightness is to minimize the chance of an LBL or IBL contaminating the KFC+SDSS-HBL sample.  Thus, we maintain with high confidence that the X-ray emission from these 39 HBLs is synchrotron emission and not SSC/EC.  Comparing FP regressions of the KFC+SDSS, KFC+SDSS-HBL, and  KFC+SDSS-LBL samples will allow us to investigate the impact of SSC/EC emission, as well as to further highlight  the importance of comparing similar regions of SEDs across the entire accreting black hole mass spectrum. 

\subsubsection{\bl\ Black Hole Masses}
\label{sec:blbhmass}
Quasar central black holes masses are normally estimated from the widths of  broad emission lines and continuum luminosities \citep[e.g.,][]{kaspi00}.  However, the featureless nature of \bl\ objects that we exploit to isolate their non-thermal jet emission also makes such measurements difficult (although see \citealt{decarli11}).  Instead, we limit ourselves to the low-redshift ($z<0.4$) \bl\ subset that show a flux component from their host galaxy in their SDSS spectra; the large size of our parent \bl\ sample ensures that a relatively large number of \bl\  objects show enough host galaxy flux.  From the measured widths of their stellar absorption lines, we infer central black hole masses using the $\mbh-\sigma$ relation \citep{tremaine02}.  We are able to measure black hole masses for 71 SDSS \bl\ objects, with a precision around 0.30-0.35~dex.  These black hole masses are available in \citet{plotkin11}, where we also describe our technique in detail.    Here, we include \nblmass\ of these 71 objects that also have X-ray detections in RASS, 39 of which we classify as HBLs (see \S \ref{sec:kfcsdsshblsamp}).   We note that our requirement of substantial host galaxy flux biases our \bl\ subset with black hole masses toward the most weakly beamed objects, thus reducing (though not  negating) the effect of Doppler boosting.  Also, since all 55 of our SDSS \bl\ objects have $z<0.4$, potential biases from luminosity evolution are  minimized.  This could be  important because \bl\ objects may have a peculiar ``negative'' cosmic evolution \citep{morris91}.

\subsubsection{\bl\ Radio Luminosities}
\label{sec:blradiolum}
\citetalias{plotkin10} explicitly consider only optical properties for their \bl\ selection.\footnote{However, multiwavelength constraints implicitly influence their selection for some sources \citep[also see][]{collinge05}.}  Post-selection, they correlated their sample to the Faint Images of the Radio Sky at Twenty cm survey \citep[FIRST;][]{becker95} and to the NRAO VLA Sky Survey \citep[NVSS;][]{condon98} radio surveys at 1.4~GHz, and to the \textit{ROSAT} All Sky Survey \citep[RASS;][]{voges99,voges00} in the X-ray (0.1-2.4~keV).  We use their published radio luminosities ($\nu L_{\nu}$) at rest-frame 5~GHz, which were estimated from FIRST/NVSS assuming a local spectral index of $\alpha_R=-0.27$.   All of the \bl\ objects considered in our FP work have firm spectroscopic redshifts from host galaxy features for estimating luminosities.  We include in our error budget 5\% uncertainties on the radio flux densities, and  5\% uncertainties on  distances (the latter is dominated by uncertainty in the Hubble flow).

\subsubsection{X-ray Luminosities}
Our goal is to better understand how the adopted method for estimating \bl\ X-ray luminosities  may influence FP regressions.  So, we estimate X-ray luminosities in three different ways to explore the effect.  First, we take each \bl\ object's observed RASS position sensitive proportional counter (PSPC) count rate from 0.1-2.4~keV.   From these count rates and each object's SDSS redshift, we estimate broadband X-ray luminosities from 0.5-10~keV rest-frame, corrected for Galactic absorption, using the Portable, Interactive Multi-Mission Simulator \citep[PIMMS;][]{mukai93}.  Hydrogen column densities along each object's sightline are taken from the \citet{stark92} hydrogen maps using the {\tt colden} tool in the \textit{Chandra} Interactive Analysis of Observations package \citep[CIAO;][]{fruscione06}.  We assume a local X-ray spectral index of $\alpha_X$=1.50 and 1.05 for HBLs and LBLs, respectively, which are typical values for each subclass \citep[see, e.g.,][]{padovani96}.  We refer to these  luminosities as ``real X-ray luminosities.'' 

Uncertainties on the ``real X-ray luminosities"  are dominated by the adopted values of $\alpha_X$, and we conservatively add an uncertainty of $\sim$25\% to each ``real X-ray luminosity" measurement.    This error budget is adopted because that is the magnitude our  ``real X-ray luminosities" would change if we instead used local X-ray spectral indices between $1<\alpha_X<3$ (which is a reasonable range observed for \bl\ objects).    We expect these ``real X-ray luminosities'' to be affected by synchrotron cooling for HBLs, and dominated by SSC/EC for LBLs (and SSC/EC potentially contributing non-negligibly to the observed X-rays for IBLs).  Thus, we do not expect FP regression coefficients using  ``real X-ray luminosities"  to match the theoretical predictions in Equation~\ref{eq:jetcoeff}.

We also estimate 0.5-10~keV rest-frame X-ray luminosities from the observed optical flux densities assuming $\alpha_X=0.6$ as in the KFC sample.  We use optical luminosity densities of the AGN component (i.e., decomposed from the host galaxy) at rest-frame 5000~\AA.   We refer to these X-ray luminosities as ``KFC-like.''  The  ``KFC-like" luminosities are sensitive to the assumed value of $\alpha_X=0.6$, since we are extrapolating these luminosities over 2-4 decades in frequency.  We thus add a factor of two uncertainty to our ``KFC-like" X-ray luminosities, the amount we expect luminosities to change if we rather choose $\alpha_X=0.5$ or $\alpha_X=0.7$ (uncertainties on distance measures are negligible compared to this factor of two uncertainty).  Note, $\alpha_X=0.6$ is a typical spectral index for optically thin synchrotron emission.  So we do not expect this choice to \textit{systematically} affect our best-fit FP (i.e.,``average") regression coefficients; rather the fact that different objects have different $\alpha_X$ values will primarily add intrinsic scatter.  If the optical waveband really is probing optically thin synchrotron emission for \bl\ objects, then we expect the ``KFC-like" luminosities to yield FP regression coefficients similar to the contracted KFC subsample.  If the \bl\ synchrotron emission is already synchrotron cooled in the optical, then we expect shallower FP coefficients (see \S\ref{sec:mcsimFRIBL}). 

Finally, we build multiwavelength SEDs for each \bl\ object by correlating to other large-scale multiwavelength surveys, including the Westerbork Northern Sky Survey (WENSS; \citealt{rengelink97}) and the Green Bank 6~cm survey (GB6; \citealt{gregory96}) in the radio, the Two-Micron All Sky Survey (2MASS; \citealt{skrutskie06}) in the near-infrared, the \textit{Galaxy Evolution Explorer} (GALEX) in the ultraviolet, and the XMM-Newton Slew Survey \citep{saxton08} in the X-rays, in addition to the FIRST/NVSS radio and RASS X-ray data points available from \citetalias{plotkin10}.   We fit parabolas (in $\log \nu F_{\nu}$ -- $\log \nu$) to these SEDs: $\log \nu F_{\nu} = A_1 (\log \nu)^2 + B_1 \log \nu + C_1$ \citep[e.g.,][]{massaro04}.  For the 39 HBLs we include X-ray data in the SED fit.  For the other 16 \bl\ objects, it is unclear from their SEDs if the observed X-ray data are probing the synchrotron or the SSC/EC component, so we omit X-ray data from the fits for those 16 objects.

From these SEDs, we measure the luminosity density at the frequency where the SED has a local spectral index $\alpha = 0.6$, corresponding to $\log \nu = 0.4/(2A_1) + \log \nu_p$, where $\log \nu_p = -B_1/2A_1$ is the peak frequency (i.e., where $d \log \nu F_{\nu}/d \log \nu = 0$).  $\log {\nu_p}$ corresponds to the synchrotron cutoff frequency.  We then extrapolate X-ray luminosities at rest-frame 0.5-10~keV from  the flux density where $\alpha=0.6$.  We refer to these X-ray luminosities as ``SED-based.''   We add a factor of two uncertainty to the ``SED-based luminosity" estimates.  We expect the FP regression with ``SED-based luminosities" to be consistent with the contracted KFC subsample regression. 

A potential advantage of the ``SED-based luminosities" over the ``KFC-like luminosities" is that the ``SED-based luminosities" are always extrapolated from optically thin synchrotron emission, regardless of the frequency of the synchrotron cutoff.    In principle, the ``KFC-like luminosities" can probe different regions of each SED from object to object (i.e., some ``KFC-like luminosities" could be synchrotron cooled),  which would affect the best-fit regression coefficients.   We thus expect that using observed SEDs to estimate ``X-ray luminosities'' (even modeling with crude parabolas) will reduce the magnitude of this effect, provided our SED models are accurate parameterizations of each \bl\ SED.

As noted at the beginning of this section, we restrict ourselves to 55 \bl\ objects with black hole mass measurements and X-ray detections in RASS (39 HBLs and 16 IBL/LBLs  for the KFC+SDSS-HBL and KFC+SDSS-LBL samples, respectively).  In principle, we could potentially include all 71 of the \citetalias{plotkin10} \bl\ objects with black hole mass measurements: the regression technique of \citetalias{kelly07} can account for upper limits when regressing the samples with ``real X-ray luminosities.''  The other two types of ``X-ray luminosities'' are extrapolated from lower frequencies, so their regressions do not require the handling of censored data.  However, comparison between the three regressions with the three different ``X-ray luminosities'' are not as uniform if some include censored data and some do not.  Perhaps more importantly, the X-ray data is very useful for constraining the HBL \bl\ SEDs and estimating accurate ``SED-based X-ray luminosities'' for HBLs.  Reducing the above systematics provides a greater benefit than adding 16 more \bl\ objects to the regression.  Thus, we choose to restrict this study exclusively to SDSS \bl\ objects with RASS X-ray detections.

\subsection{Bayesian Regression Including SDSS BL Lac Objects}
\label{sec:bayessdss}
\begin{figure}
 \includegraphics[scale=0.93]{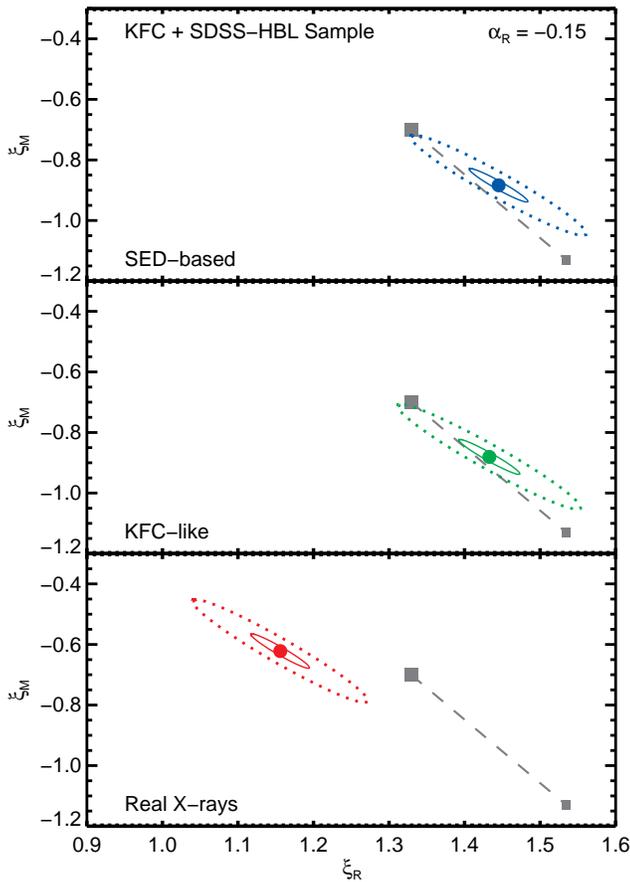}
 \caption{Best-fit coefficients from the Bayesian regression of the KFC+SDSS-HBL sample.  For clarity,  we do not show the best-fit coefficients for the KFC+SDSS or the KFC+SDSS-LBL samples (but see Table~\ref{tab:sdssbayes}.)   From top to bottom we show the best-fit regression coefficients when using SED-based (blue circle), KFC-like (green circle), and real X-ray luminosities (red circle) for the \bl\ objects, respectively, with 1$\sigma$ (solid line)  and 3$\sigma$ (dotted line) error ellipses overdrawn.  As in previous figures, the filled squares connected with a dashed line are the theoretical FP coefficient predictions for the jet model, for $\alpha_R=-0.15$ and $2<p<2.8$, as relevant to the KFC+SDSS samples.  This figure appears in color in the online version of this article. }
 \label{fig:dr7bayes}
 \end{figure}

 The best-fit FP coefficients for the KFC+SDSS, KFC+SDSS-HBL, and KFC+SDSS-LBL samples are shown in Table~\ref{tab:sdssbayes}, and the results for the KFC+SDSS-HBL sample are illustrated in Figure~\ref{fig:dr7bayes}.  Using real X-ray data clearly gives FP slopes shallower than any predictions from the optically thin synchrotron model.  Thus, X-ray luminosities indeed need to be extrapolated from lower frequencies for the most massive black holes, if their SEDs are jet dominated.  The effect of   SSC/EC emission contributing to the observed X-rays from LBLs  is evident by comparing the best-fit ``real X-ray" coefficients for the KFC+SDSS, KFC+SDSS-HBL, and the KFC+SDSS-LBL samples.  The inclusion of  LBLs yields shallower coefficients.

 \begin{table*}
\caption{Bayesian Regression of the KFC+SDSS Samples}
\label{tab:sdssbayes}
\scriptsize
\begin{tabular}{l  cc r@{.}l c cc r@{.}l c }
\hline
		  Type of X-ray                  & %1
		 %                    & %2
		 \multicolumn{5}{c}{Beamed \bl\ Objects}   & %3-6 
		 \multicolumn{5}{c}{Debeamed \bl\ Objects}  \\         
		   \cline{2-6}       \cline{7-11} 
		Luminosity      &  %1
	        % Number           &  %2           
	         	$\xi_R$  	   &   %3
		$\xi_M$     &  %4
		\multicolumn{2}{c}{$B$}             & %5
		$\sigma_{int}$ & %6
		$\xi_R$      & %7
		$\xi_M$      &	%8
		\multicolumn{2}{c}{$B$}            & %9
		$\sigma_{int}$ \\ %10
		\hline
		
		\multicolumn{11}{l}{ \underline{KFC+SDSS-HBL Sample (82 Objects)} } \\ 
	        SED-based &        1.45 $\pm$ 0.04 &  $-$0.88 $\pm$ 0.06 &  $-$6&07 $\pm$ 1.10 &   0.07 $\pm$ 0.05 &   1.42 $\pm$ 0.05 &  $-$0.87 $\pm$ 0.07 &  $-$5&30 $\pm$ 1.52 &  0.07 $\pm$ 0.04 \\ 
	         KFC-like &           1.43 $\pm$ 0.04 &  $-$0.88 $\pm$ 0.06 &  $-$5&71 $\pm$ 1.14 &   0.08 $\pm$ 0.04 &   1.41 $\pm$ 0.04 &  $-$0.86 $\pm$ 0.06 &  $-$4&90 $\pm$ 1.28 &  0.08 $\pm$ 0.04 \\ 
                Real X-ray &  1.16 $\pm$ 0.04 &  $-$0.62 $\pm$ 0.06 &   2&27 $\pm$ 1.10 &   0.45 $\pm$ 0.04 &   0.75 $\pm$ 0.07 &  $-$0.25 $\pm$ 0.08 &  13&85 $\pm$ 1.88 &  0.56 $\pm$ 0.05 \\ 
                
                \hline
                \multicolumn{11}{l}{\underline{KFC+SDSS-LBL Sample (59 Objects)}} \\
                 SED-based &        1.47  $\pm$ 0.04 &  $-$0.92 $\pm$ 0.06 &  $-$6&91 $\pm$ 1.19 &   0.16 $\pm$ 0.08 &   1.46 $\pm$ 0.05&  $-$0.90 $\pm$ 0.07 &  $-$6&47 $\pm$ 1.50 &  0.16 $\pm$ 0.08 \\ 
	         KFC-like &           1.35 $\pm$ 0.04 &  $-$0.81 $\pm$ 0.05 &  $-$3&24 $\pm$ 1.05 &   0.12  $\pm$ 0.07 &   1.32 $\pm$ 0.05 &  $-$0.79 $\pm$ 0.06 &  $-$2&30 $\pm$ 1.43 &  0.13 $\pm$ 0.07 \\ 
                Real X-ray &  0.96 $\pm$ 0.04 &  $-$0.43 $\pm$ 0.06 &   7&91 $\pm$ 1.23 &   0.52 $\pm$ 0.05 &   0.67 $\pm$ 0.07 &  $-$0.17 $\pm$ 0.09 &  16&13 $\pm$ 2.09 &  0.65 $\pm$ 0.07 \\ 
                
                \hline 
                \multicolumn{11}{l}{ \underline{KFC+SDSS Sample (98 Objects)} } \\
                	SED-based &        1.45 $\pm$ 0.04 &  $-$0.89 $\pm$ 0.06 &  $-$6&32 $\pm$ 1.13 &   0.11 $\pm$ 0.06 &   1.44 $\pm$ 0.05 &  $-$0.88 $\pm$ 0.07 &  $-$5&80 $\pm$ 1.45 &  0.11 $\pm$ 0.05 \\ 
	         KFC-like &           1.41 $\pm$ 0.03 &  $-$0.86 $\pm$ 0.05 &  $-$4&92 $\pm$ 0.99 &   0.07 $\pm$ 0.04 &   1.37 $\pm$ 0.05 &  $-$0.82 $\pm$ 0.06 &  $-$3&77 $\pm$ 1.37 &  0.07 $\pm$ 0.04 \\ 
                Real X-ray &  1.06 $\pm$ 0.04 &  $-$0.51 $\pm$ 0.06 &   5&08 $\pm$ 1.14 &   0.55 $\pm$ 0.04 &   0.68 $\pm$ 0.06 &  $-$0.16 $\pm$ 0.08 &  16&04 $\pm$ 1.69 &  0.56 $\pm$ 0.04 \\ 
                 \hline
\end{tabular}
\end{table*}

Interestingly, the best-fit ``real X-ray'' coefficients for the KFC+SDSS-HBL regression are nearly identical to the Bayesian regression coefficients for the contracted+FR~I KFC subsample.  Both samples therefore  suffer from a similar type of bias due to the most massive black holes that is of comparable magnitude.  Because we are certain that  HBL X-rays are predominantly synchrotron cooled radiation (i.e., there is little to no contamination from SSC/EC or accretion flow X-rays), emission in the optical waveband from the KFC FR~I galaxies is also likely synchrotron cooled.  Our FP simulations in \S \ref{sec:mcsimFRIBL} further support the interpretation that FR~I galaxies emit synchrotron cooled radiation in the optical.  Both sets of coefficients are  consistent with the simulation  where we modeled X-rays from lower mass black holes dominated by optically thin synchrotron ($\alpha_X=0.6$) but  X-rays from the most massive black holes ($>\approx 10^8~\msun$) are synchrotron cooled ($\alpha_X=1.0$; see red cross in Figure~\ref{fig:mcsimDiffp}).

\citet{heinz04} show how to incorporate synchrotron cooling into the equations for scale-invariant jets by adding another scale length to the problem -- the scaled-distance from the black hole where electrons become synchrotron cooled.   The X-ray luminosity for a synchrotron cooled, scale-invariant jet should then follow $L_X \propto M\dot{m}$, i.e., similar to that for a radiatively efficient ($q=1$)  standard accretion disk.\footnote{Optically thin X-ray jet emission would follow $L_X \propto (M\dot{m})^{(17/12) + (2/3)\alpha_X}$ (\citealt{markoff03}; \citetalias{falcke04}).} %%
Cooling affects the dynamics of only the highest energy relativistic electrons, so the expected radio emission remains unchanged.  \citet{heinz04} then derive the expected FP coefficients for synchrotron cooled X-rays, and finds that one expects shallower slopes than if the X-rays are optically thin.  Thus, our regression with ``real X-rays'' is biased toward the expected direction.  However, it is not possible to directly compare our best-fit coefficients to the prediction from \citet{heinz04} more rigorously because the KFC X-ray luminosities for lower-mass black holes (i.e., GBHs, SgrA$^{\star}$, and LLAGN) are probing optically thin (i.e., uncooled) synchrotron in the context of the jet model.

The ``SED-based" regression coefficients for  the KFC+SDSS, KFC+SDSS-HBL, and KFC+SDSS-LBL samples are all consistent (within $\pm$1$\sigma$) with the regression coefficients for the contracted KFC subsample.  We thus conclude that SED modeling allows accurate measurements for optically thin synchrotron luminosities.  We also find that the ``KFC-like" regression of KFC+SDSS-HBL sample is  consistent with the contracted KFC subsample.  Thus, unlike for FR~I galaxies, it is generally appropriate to extrapolate HBL ``X-ray luminosities'' from  the optical.  That is, because \bl\ jet emission is Doppler boosted, the optical waveband typically probes \textit{optically thin} synchrotron emission for HBLs. However, the ``KFC-like" regression of the KFC+SDSS-LBL sample is \textit{not} consistent with the contracted KFC subsample.  We argue below that, like FR~I galaxies, the discrepancy is because LBL synchrotron emission (for at least some LBLs) is already synchrotron cooled in the optical waveband.\footnote{The ``KFC-like" regression coefficients for the KFC+SDSS sample naturally lie in between those for the KFC+SDSS-HBL and KFC+SDSS-LBL samples (since the latter two samples are subsets of the KFC+SDSS sample).}

\begin{figure}
 \includegraphics[scale=0.88]{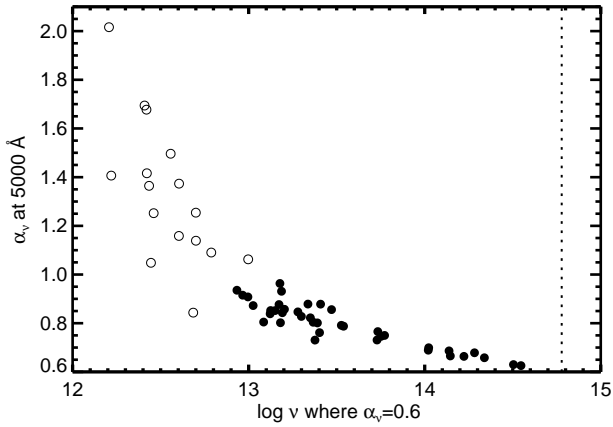}
 \caption{The local spectral index ($\alpha_\nu$) of each SED fit at rest-frame 5000~\AA\ vs.\ $\log$ frequency where each SED fit has $\alpha_\nu=0.6$.   The filled circles are HBLs ($\alpha_{rx}<0.7$), and the open circles are IBL/LBLs ($\alpha_{rx}>0.7$). The dotted vertical line marks 5000~\AA\ for reference.  All HBLs have spectral indices at 5000~\AA\ indicative of optically thin synchrotron emission.  However,  synchrotron cooling may mildly bias the regression toward shallower best-fit coefficients if LBL/IBL X-ray luminosities are extrapolated from the optical.  This effect is much more significant for FR~I galaxies. }
 \label{fig:blcooling}
 \end{figure}

In Figure~\ref{fig:blcooling} we show the spectral index at rest-frame 5000~\AA\ (i.e., where we measure optical luminosities for the ``KFC-like'' regression) inferred from our SED fits, and we compare to the frequency where each SED has $\alpha_\nu=0.6$ (i.e., from which we measure ``SED-based" luminosities).  All of the HBLs (filled circles) have spectral indices indicative of optically thin synchrotron emission ($0.5 < \alpha_{\nu}<1.0$) at rest-frame 5000~\AA.\footnote{All of the HBLs also have synchrotron peak (i.e., cutoff) frequencies $\nu_p > 10^{14.9}$~Hz, consistent with the expected cutoff frequencies for HBLs \citep[e.g., see \S6 of][]{abdo10_sed}.} Thus, the ``KFC-like'' and ``SED-based" luminosities probe similar regions of HBL SEDs; so, we expect the FP regression coefficients  of the KFC+SDSS-HBL sample using those two luminosity estimates to then be  similar and consistent with the contracted KFC subsample regression (and indeed they are almost indistinguishable).

On the other hand,  IBL/LBLs (open circles) have steeper spectral indices at rest-frame 5000~\AA, and many LBLs even appear to be radiatively cooled ($\alpha_{\nu}>1$).  Therefore, the KFC+SDSS-LBL regression coefficients should be shallower than the KFC+SDSS-HBL coefficients.   Thus, if one has the requisite data available to model SEDs, then that is a preferable method for estimating ``X-ray luminosities" from very massive accreting black holes with jet-dominated SEDs.   Note, the ``KFC-like" coefficients for the KFC+SDSS-LBL sample are not as shallow as the ``real X-ray" coefficients for the KFC+SDSS-HBL sample because apparently not all IBL/LBLs are very strongly affected by synchrotron cooling at 5000~\AA\ rest-frame.

From Figure~\ref{fig:blcooling}, we can expand further on  why FR~I ``X-ray luminosities" cannot simply be extrapolated from the optical. For unbeamed AGN with the most massive SMBHs (i.e., FR~Is), the jet will appear to become optically thin at  much lower frequencies (by almost an order of magnitude if \bl\ objects have Doppler parameters $\delta\sim7$; also see Figure~6 of \citealt{balmaverde06} for a sketch of how FR~I and \bl\ SEDs differ because of  Doppler beaming).  From Figure~\ref{fig:blcooling}, we estimate a debeamed \bl\ object would have a steeper $\alpha_\nu$ at 5000~\AA\ so that $\alpha_\nu > 0.8$ always (and most with $\alpha_\nu > 1.0$).   Thus, optical nuclear luminosities of FR~Is should strongly be affected by synchrotron cooling.    SED-modeling of unbeamed jet-dominated AGN with very massive central black holes is thus necessary to place them onto the FP.

 The KFC+SDSS-HBL sample minimizes  concern of synchrotron cooling systematically biasing the FP regression.   We thus consider the following regression to be the most robust:
 \begin{eqnarray}
\nonumber \log L_x  & =& (1.45 \pm 0.04) \log L_R - (0.88 \pm 0.06) \log \mbh \\
                                      &    & - 6.07 \pm 1.10,
\label{eq:beamedFP}
\end{eqnarray} 
\noindent  To our knowledge, because of our sample selection and adopted regression technique,  Equation~\ref{eq:beamedFP} is the most accurate FP regression to date  for sub-Eddington accreting black holes with flat/inverted radio spectra. 
 
 For illustrative purposes, and comparison to previous FP studies, we  show a projection of our final FP in Figure~\ref{fig:fp}.   Shown is the best-fit for the KFC+SDSS-HBL sample, with ``SED-based'' X-ray luminosities, both observed (top panel) and corrected for Doppler beaming (bottom panel; see \S\ref{sec:dr7blDoppler}).  For reference, we also show the location of FR~I galaxies on the FP (with ``X-ray luminosities" extrapolated from the optical), although they are not included in the fit.  As expected,  FR~I galaxies tend to undershoot the FP.  We note that regressing $L_X$ and $L_R$ just for the SDSS~\bl\ objects does not follow the same slope as the FP.  This result is due to the limited dynamic range when considering only the \bl\ objects, and the relatively large intrinsic scatter.  Also,  the dispersion in our SDSS \bl\ black hole mass measurements are of the same order as the measurement errors, $\sim$0.30-0.35~dex.  Thus, we do not expect to be able to infer reliable FP coefficients, especially for $\xi_M$,  when considering only \bl\ objects.
   
\subsubsection{Doppler Boosting}
\label{sec:dr7blDoppler}
\begin{figure*}
 \includegraphics[scale=1.0]{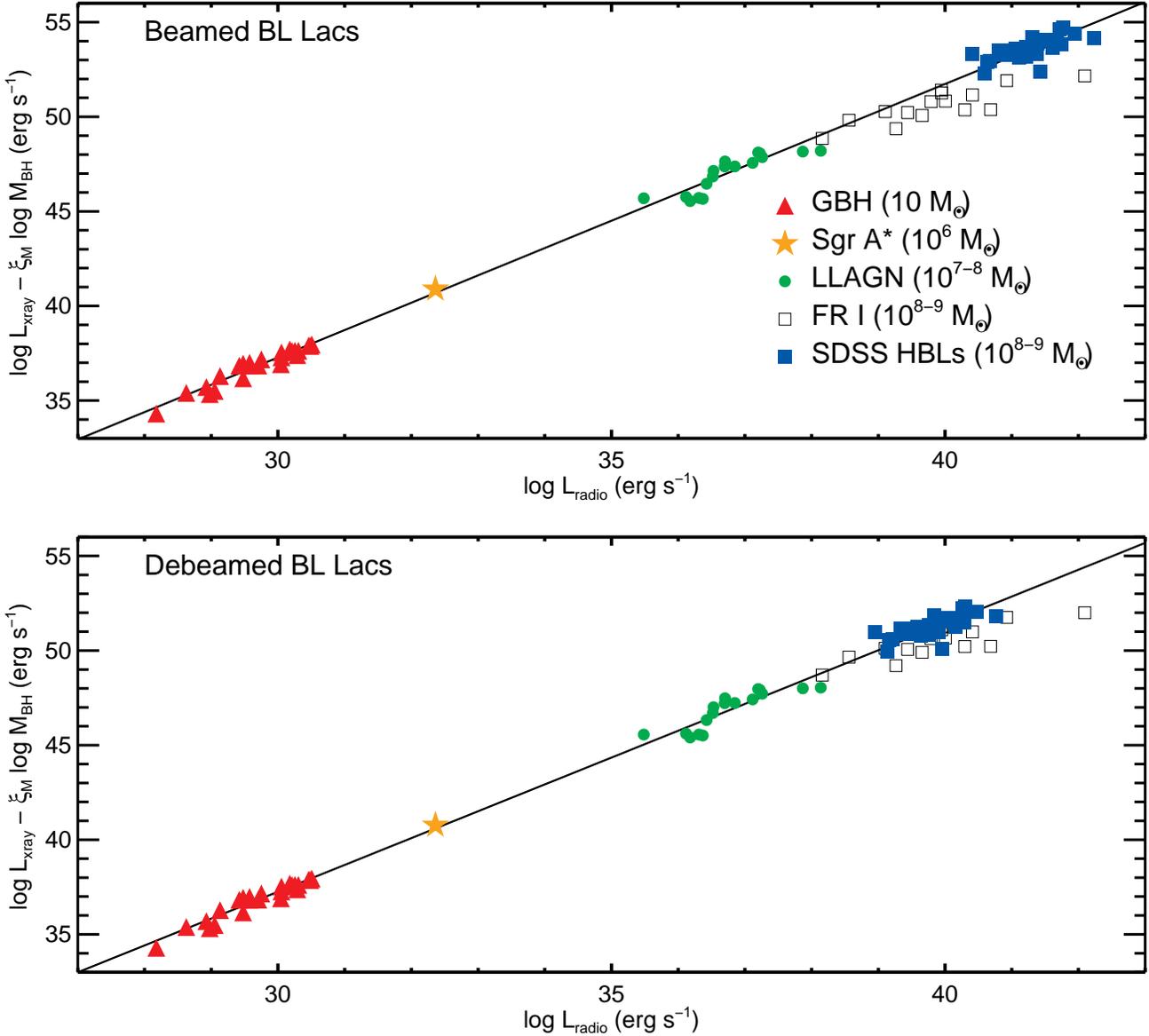}
 \caption{Our best-fit FP for low-accretion rate black holes (the KFC+SDSS-HBL sample; 82 objects) using the Bayesian regression algorithm and SED-based X-ray luminosities.  The top panel shows the regression for beamed \bl\ objects, and the \bl\ objects are debeamed in the bottom panel.  FR~I galaxies are shown for reference, but they are not included in the  regression.  This figure appears in color in the online version of this article.}
 \label{fig:fp}
 \end{figure*}

Throughout, we have assumed that Doppler beaming only affects \bl\ objects, and that, to first order, Doppler beaming does not largely affect the best-fit coefficients.  That is, we assume that optically thick and optically thin synchrotron emission are beamed by approximately the same amount, so beaming primarily only moves \bl\ objects up or down the best-fit line in Figure~\ref{fig:fp} \citep[also see][]{falcke04, heinz04_beaming, li08}.   We  note that \citet{landt02} find no \textit{statistical} evidence that more weakly beamed \bl\ objects (like the 55 \bl\ objects used in this work) have significantly different Doppler factors in the radio and X-ray.  Still, our assumption that the Doppler factor is similar in different wavebands is unlikely to be strictly true, and different Doppler factors probably contribute to some of the observed scatter.  

We test the effect of Doppler boosting by debeaming the SDSS \bl\ objects, assuming $\delta=7$ \citepalias{falcke04}, and thus reducing their radio and ``X-ray luminosities'' by a factor $\delta^{2+\alpha_\nu}$ \citep{lind85, urry95}.   We assume $\alpha_{\nu}=-0.27$ in the radio, $\alpha_{\nu}=0.6$ for KFC-like and SED-based X-ray luminosities, and $\alpha_{\nu} = 1.5$ and $1.05$ for HBL and LBL ``real X-ray luminosities", respectively.   The debeamed \bl\ objects have similar luminosities as FR~I galaxies, as expected from the standard AGN unification paradigm (see Figure~\ref{fig:fp}).

We then regress the KFC+SDSS, KFC+SDSS-HBL, and KFC+SDS-LBL samples with the debeamed \bl\ luminosities.   As long as our ``X-ray luminosities" are probing optically thin synchrotron radiation, debeaming the \bl\ luminosities does not significantly change the best-fit regression coefficients (see the coefficients for all three ``SED-based" regressions and the ``KFC-like" regression for the KFC+SDSS-HBL sample in Table~\ref{tab:sdssbayes}).
 
We thus conclude that Doppler beaming does not strongly influence the FP regression.  Other systematics, such as making sure one probes consistent regions of accreting black hole SEDs across the mass scale, are likely more important.
 For example, when our ``X-ray luminosities'' instead primarily probe synchrotron cooled emission or SSC/EC, as for the ``real X-ray" regressions,  then debeaming the \bl\ objects does change the best-fit FP slopes.    However, these regressions are not physically meaningful because we are probing different SED regions for different black hole masses.    Debeaming synchrotron cooled``X-ray luminosities" would move  \bl\ objects down the synchrotron cooled radio/X-ray/mass correlation predicted by \citet{heinz04}, and not the ``optically thin synchrotron" correlation followed by the lower-mass contracted KFC subsample.   Thus, debeamed synchrotron cooled or SSC/EC X-rays will bias the best-fit FP regression by a different amount than beamed  X-rays.

\section{Discussion}
\label{sec:disc}

\subsection{Comparison of the Discovery Papers -- Two Fundamental Planes?}
\label{sec:merloniVsFalcke}
A few more comments on the two discovery papers and their inferred  mechanisms for X-ray emission are in order.    Because \citetalias{merloni03} favors a RIAF and \citetalias{falcke04} a jet, there is sometimes an understandable but unfair perception in the literature that the two papers  contradict each other.  However,  \citetalias{merloni03} also outline how jet X-rays can lead to the FP, and in that case they predict coefficients similar to that of \citetalias{falcke04}.    There are likely  at least two FPs\footnote{In fact, we have already alluded to a ``third FP'' in this paper from \citet{heinz04}  if synchrotron cooled X-rays are considered instead of optically thin jet synchrotron.}, %%
and regressions of the FP only test the ``average'' emission mechanism.   The situation is actually even more complicated, in that every source follows its own FP to some level due to different values of $\alpha_R$, $p$, and $q$, which contributes to the observed intrinsic scatter about the FP. 

The apparently discrepant conclusions in the two discovery papers are an artifact of  the samples used in each paper to test the FP.   \citetalias{falcke04} considered only low-accretion rate (i.e., jet-dominated) sources, while \citetalias{merloni03} assembled a more diverse sample; the \citetalias{merloni03} sample mixes some jet and some RIAF X-ray dominated black holes, which naturally increases the intrinsic scatter.   However, \citetalias{merloni03} exclude \bl\ objects, so their most-massive SMBHs include almost exclusively more luminous quasars that are well accepted to have coronal dominated X-ray emission.  Given that those AGN provide the largest dynamic range and considering the previous section's results, it is  not surprising that the M03 regression favors the RIAF models (albeit with large intrinsic scatter).  That the FP coefficients are sensitive to the sample adopted in the two discovery papers (especially lower accretion rate objects seeming to favor jet X-rays), supports the idea that AGN mimic accretion states of GBHs.

 \citetalias{koerding06}'s reanalysis of the \citetalias{merloni03} and \citetalias{falcke04} samples seems to confirm the conclusions of the discovery papers.  However, the coefficients found by \citetalias{koerding06} for the two samples (using the same Merit function technique) are not remarkably different when considering the error bars; for the \citetalias{merloni03} sample, $\left(\xi_R\right)_{M03} = 1.45\pm0.17$ and $\left(\xi_M\right)_{M03}=-0.99\pm0.22$; for the \citetalias{falcke04} sample, $\left(\xi_R\right)_{F04}=1.41\pm0.11$ and $\left(\xi_M\right)_{F04}=-0.87\pm0.14$.  It is thus an interesting ``conspiracy'' (as \citetalias{koerding06} call it) that the two emission mechanisms yield FPs with so much overlap.  This is in large part due to optically thin jet synchrotron emission and RIAF X-ray emission both approximately following $L_X\sim M \dot{m}^2$ \citep{falcke95, markoff03}.  Statistical techniques more sophisticated than the Merit function are necessary to regress the FP with high enough accuracy to distinguish between the two radiation mechanisms.

\subsection{Comments on Other FP Studies}
\subsubsection{\citet{yuan09}}
There is a well-established global radio/X-ray correlation followed by quiescent (i.e., hard state) GBHs: $\log L_R \sim 0.7 \log L_X$ \citep{corbel03,gallo03}.  \citet{yuan09} claim this correlation is best explained if the X-rays are predominantly from the accretion flow.  This claim is based on  \citet{yuan05}, who  model emission from accreting black holes with components from both a jet and from an ADAF.  Their ADAF  accounts for the effects of outflows and convective instabilities, and it includes emission from synchrotron and bremsstrahlung processes and their thermal Comptonization.  \citet{yuan05} argue that only when the X-ray luminosity drops below a critical X-ray luminosity, $L_{X,crit} \sim10^{-5}$-- $10^{-6}~L_{Edd}$, can jet synchrotron  instead dominate the observed X-ray emission.  They predict that the  $\log L_R \sim  0.7 \log L_X$ correlation will then steepen to $\log L_R \sim 1.23 \log L_X$, as the X-ray emission switches from being ADAF to jet dominated.\footnote{Note, different from the convention in this paper, \citet{yuan05} and \citet{yuan09} cast $L_R$ as their dependent variable.  So, their prediction for a steeper correlation  at lower X-ray luminosities corresponds to a shallower (i.e., lower) value of $\xi_R$ in our notation using $\log L_X$ as the dependent variable.}   
 
   \citet{yuan09} use a sample of  22 very low-accretion rate AGN with $L_X < 10^{-6}~L_{Edd}$ (i.e., below $L_{X,crit}$) to regress the FP and test the prediction of \citet{yuan05}.  They find $\log L_R \sim (1.22 \pm 0.02) \log L_X + (0.23 \pm 0.03) \log \mbh$, in excellent agreement with \citet{yuan05}.\footnote{It is surprising that \citet{yuan05} report uncertainties on their coefficients smaller than both  our work and \citetalias{merloni03}, since the latter two studies adopt much larger samples (plus we use an arguably more accurate regression method).  Comparing to \citet{gultekin09}, who used a similar regression technique on a similarly sized sample as \citet{yuan09}, \citet{yuan09} seem to underestimate errors on their coefficients by an order of magnitude.}  %%
\citet{yuan09} do not include any AGN with $L_X \geq L_{X,crit}$, but they note that the accreting black holes in the sample used by \citetalias{merloni03} all have $L_X>L_{X,crit}$.   \citetalias{merloni03} find a shallower FP regression for their more luminous accreting black holes [$\log L_R\sim (0.60^{+0.11}_{-0.11}) \log L_X + (0.78^{+0.11}_{-0.09}) \log \mbh$], which is  consistent with the radio/X-ray correlation for quiescent GBHs (i.e., $\log L_R \sim 0.7 \log L_X$; \citealt{gallo03}).  The conclusion of \citet{yuan09} therefore supports \citet{yuan05}, that the radio/X-ray correlation  of quiescent black holes described in \citet{gallo03} is due to the jet in the radio and inverse Compton in the X-ray.  However, below a critical luminosity, the radio/X-ray correlation steepens as the X-ray emission instead becomes dominated by  jet synchrotron.
 
The argument in \citet{yuan05} for a steeper correlation when $L_X< L_{X,crit}$ assumes that  jet X-ray emission scales \textit{linearly} with $\dot{M}$.  Then, $L_{X,crit}$ corresponds to the critical  accretion rate  where jet X-rays start to outshine the accretion flow  (RIAF X-ray emission  scales approximately \textit{quadaratically} with $\dot{M}$).  Their assumed jet X-ray dependence, $L_X \propto \dot{M}$, implies that the jet X-rays are synchrotron cooled \citep[see][]{heinz04}.   On the other hand, if jet emission is still optically thin in the X-rays for GBHs, then $L_X\sim \dot{M}^{(17/12) + (2/3)\alpha_X}$ (\citealt{markoff03}; \citetalias{falcke04}).   Then, optically thin synchrotron can dominate over coronal emission in the X-rays at higher values of $L_X/L_{Edd}$, and the \citet{gallo03} correlation can be interpreted as predominantly optically thin jet synchrotron X-rays.  There is  evidence that the X-ray flux of the GBH XTE J1550-564 is 100\% optically thin jet synchrotron at $L_X \sim 4 \times 10^{-5} - 4 \times 10^{-4}~L_{Edd}$ as it fades back into the low-hard state following an outburst \citep{russell10}.\footnote{We convert the bolometric luminosities quoted in \citet{russell10} to that in the 2-10~keV X-ray band.}  %%
This X-ray luminosity is higher, although not totally inconsistent, with the value of $L_{X,crit}$ predicted by \citet{yuan05}.  However, an important difference is that \citet{russell10} measure an X-ray spectral index $\alpha_X\sim0.7$, indicating optically thin (i.e., uncooled) X-ray emission accounts for 100\% of the X-ray flux of XTE J1550-564 at energies of a few~keV (i.e. X-ray luminosity should scale approximately quadratically with $\dot{M}$ and not linearly, inconsistent with the assumption of \citealt{yuan05}).   It is also important to note that \citet{russell10} find synchrotron emission cannot contribute all the hard X-rays at the beginning of the outburst, so at least one other parameter besides accretion rate is likely important for controlling the X-ray emission mechanism.   

\citet{russell10}  note that the synchrotron cutoff is believed to occur at $>$40--100~keV for hard-state GBHs, above the observed hard X-ray band (see their \S 4.1).  We might then expect, if the quiescent GBHs used to measure the \citet{gallo03} radio/X-ray correlation were observed at X-ray energies above the cutoff, then they would follow a correlation closer to the $\log L_R \sim 1.23 \log L_X$ correlation predicted by \citet{yuan05}.  That is, the change in slope does not necessarily require a switch between RIAF and jet dominated X-rays.  Rather, it could instead more heavily depend on the waveband in which ``X-ray luminosities'' are measured.

We argue that it is this type of observational effect responsible for the different  FP slopes measured by \citet{yuan09} compared to \citetalias{merloni03}.  More precisely, the X-ray observations of the AGN sample used by \citet{yuan09} are affected by synchrotron cooling, while very few AGN  in the sample used by \citetalias{merloni03} are emitting predominantly synchrotron cooled X-rays.   \citet{yuan09} use the lowest-luminosity AGN yet in any FP study, but most have such low $L_{X}/L_{Edd}$ because they have relatively large black hole masses.  Of their 22 AGN, 14  have $M>10^8$~$\msun$.  Thus, if their X-rays are indeed due to jet emission, they will be synchrotron cooled, and lower best-fit values of $\xi_R$ (as predicted by \citealt{heinz04} and \citealt{yuan05}) are expected.  

The premise behind \citet{yuan09} is valid, that synchrotron cooled X-ray emission can outshine the corona if $L_X/L_{Edd}$ is low-enough, and indeed their 22 AGN seem to have X-rays dominated by cooled synchrotron emission.  However,  comparing their results to the \citetalias{merloni03} coefficients is not evidence that  \textit{all} accreting black holes with $L_X > L_{X,crit}$ have coronal dominated X-ray emission.  \citetalias{merloni03} omit  $L_X > L_{X,crit}$ AGN with jet-dominated SEDs (i.e., \bl\ objects) from their sample, and this influences their regression toward FP coefficients predicted by the RIAF model used by \citetalias{merloni03}.  Based on our results in \S \ref{sec:bayessdss}, if \citetalias{merloni03} had included \bl\ objects (with $L_X > L_{X,crit}$) and used ``real X-ray luminosities'' for them,  then \citetalias{merloni03} also would have recovered coefficients more similar to those recovered by \citet{yuan09} (e.g., Figure~\ref{fig:dr7bayes}).   \citetalias{merloni03} note their FP scalings are not valid if synchrotron cooling becomes important, so one must be very cautious comparing the low-luminosity AGN in \citet{yuan09} to \citetalias{merloni03}, as such a comparison is not uniform.

\citet{degasperin11} report on sensitive radio and X-ray observations of 16 Type 2 LLAGN from the SDSS.\footnote{Their sample actually includes a 17$^{th}$ AGN, but they note that that source is most likely a \bl\ object and we thus exclude it from this discussion.} %%
  Their LLAGN sample covers a large dynamic range in luminosity, but the sample is selected to cover a narrow range in black hole mass ($10^8 < \mbh < 10^{8.5}~\msun$).   From their radio, optical, and X-ray data, they conclude that these LLAGN have radio and X-ray emission dominated by jet radiation.   \citet{degasperin11} also use their sample to regress the FP, and they find that the FP coefficients of their sample are consistent with that of \citet{yuan09} (and inconsistent with both \citetalias{merloni03} and \citetalias{koerding06}).   However, all of the LLAGN in the sample of \citet{degasperin11} have X-ray luminosities brighter (most by  more than an order of magnitude) than the value of $L_{X,crit}$ predicted by \citet{yuan05}.  These higher luminosities make it challenging to interpret the consistency with \citet{yuan09} being due to a transition from corona to jet dominated X-ray emission below a critical X-ray luminosity.  Again, we argue that such an interpretation is not necessary.  The LLAGN in the \citet{degasperin11} sample have relatively massive central black holes, and their X-ray emission is very likely affected by synchrotron cooling.   Thus, the  FP slopes recovered by  \citet{degasperin11} are naturally explained if, because of their very massive central black holes, their X-ray observations (from 2-10~keV) are probing primarily synchrotron cooled jet X-ray emission (opposed to optically thin synchrotron that would be probed by  X-ray observations if the AGN had lower central black hole masses).  

 \citet{gallo06}  obtained simultaneous radio and X-ray observations for the GBH A0620-00 in quiescence with extremely low $L_X=10^{-8.5}~L_{Edd}$.   They show that the observed radio luminosity of A0620-00 is consistent with extrapolating the $\log L_R \sim 0.7 \log L_X$ correlation to low $L_X$, inconsistent with the \citet{yuan05} prediction that the correlation should steepen below $L_{X,crit}=10^{-5} - 10^{-6}~L_{Edd}$.   It is true that the GBH radio--X-ray luminosity correlation has some intrinsic scatter, and one  cannot definitively extend to very low accretion rates based on only one data point.  However, if the correlation steepens to $L_R\sim L_X^{1.23}$ at $L_{X,crit}<10^{-5.5}~L_{Edd}$, then  A0620-00 should be radio-fainter than the  $\log L_R \sim 0.7 \log L_X$ correlation by a factor of 39.  This is larger than even a generous $\sigma_{int}=1.00$~dex scatter in radio luminosity.  
   
Finally, FR~I and II radio galaxies may yield additional insight.  X-ray emission in FR~I galaxies is likely dominated by the jet, while more luminous FR~II galaxies primarily emit coronal X-rays (see \S \ref{sec:whyfribad}, and references therein).  The so-called FR~I/II dichotomy \citep[e.g.,][]{ledlow96} could be due to a state transition analogous to that of hard and soft-state GBHs; \citet{wold07} argue the state transition to jet dominated X-rays occurs near $\dot{m}=0.004$ and is a factor of 10 lower for FR~I/IIs than for GBHs.  Estimates of a state transition in blazars (i.e., between \bl\ objects and flat spectrum radio quasars) are similarly placed around 0.01$L_{Edd}$ \citep[e.g.,][]{ghisellini09}. Scaling downward, then we speculate it is reasonable to think sub-Eddington GBHs could also show jet dominated SEDs below a few percent $L_{Edd}$.  While further study is needed, we conclude based on all the evidence presented in this subsection, it is very likely that some accreting stellar mass black holes with $L_X>10^{-5} - 10^{-6}~L_{Edd}$ can have jet dominated SEDs in the X-ray waveband.  

\subsubsection{\citet{li08}}
\label{sec:li08}
\citet{li08} examined the FP by regressing a subset of X-ray selected broad-line AGN (i.e., excluding \bl\ objects) from the SDSS and RASS (their AGN are taken from the \citealt{anderson07} X-ray AGN catalog).\footnote{\citet{li08} is an updated version of \citet{wang06} to include a larger X-ray selected AGN sample.  \citet{wang06} used AGN from  a similar, but smaller,  X-ray selected SDSS AGN sample from \citet{anderson03}}  %%
Of the 10$^4$ spectroscopically confirmed X-ray emitting AGN in \citet{anderson07}, \citet{li08} use 725 in their study for which they can measure black hole masses from widths of broad emission lines.   Their sample is uniform, and to our knowledge,  the largest number of objects implemented in a FP study to date.  \citet{li08}  find negligible dependence on black hole mass, which they attribute to using different rest-frame radio and X-ray luminosities than \citetalias{merloni03} (as well as to the heterogeneous nature of the black hole mass measurements adopted by \citetalias{merloni03}).   A perhaps more  important consideration is that  \citet{li08} only use SDSS AGN with large black hole masses in their regression (each with relatively large uncertainties).   There is thus unlikely  a large enough dynamic range in black hole mass to reliably recover $\xi_M$.  From their Figure~1, their black hole masses are peaked near 10$^{8.5}$--10$^{9.0}$~$\msun$; also see \citealt{kelly07_bhmass} on estimating AGN central black hole masses from single epoch spectroscopy of broad emission lines, and how the inferred distribution of mass measurements may appear broader than in reality.  

 \citet{anderson07} also include $\sim$250 \bl\ objects in their  AGN catalog, which are excluded from the \citet{li08} subsample since they do not display strong enough emission lines to measure black hole masses.  The \bl\ selection algorithms implemented in \citetalias{plotkin10}, our parent sample of \bl\ objects in this work, were largely based on those developed in \citet{anderson07} (combined with \citealt{anderson03}).    Thus, the unbeamed AGN used by \citet{li08} cover similar dynamic ranges in luminosity and black hole mass, with similarly sized error bars, as the beamed AGN used in our work.  We find it is not possible to obtain reliable coefficients considering only SDSS \bl\ objects.  Rather, massive accreting black holes should be used in conjunction with their lower-mass counterparts.

Finally, we note an interesting result that \citet{li08} find different correlations for radio-loud and radio-quiet SDSS AGN ($L_R\propto L_X^{1.4}$ if radio-loud, and $L_R\propto L_X^{0.7}$ if radio-quiet).  Given the large black hole masses of the sample used by \citet{li08}, their radio-loud AGN X-ray luminosities (as observed by RASS) would be affected by synchrotron cooling if jet dominated.   The different correlations may then be consistent with their radio-loud AGN emitting primiarly synchrotron cooled jet emission in the X-rays, while the RIAF is predominantly responsible for the radio-quiet objects' X-ray emission.    First,  however, further study into potential selection biases is necessary.  For example, due to the \citet{anderson07} AGN being X-ray selected, perhaps the requirement of X-ray emission from RASS preferentially excludes  radio-loud AGN at low-radio flux densities.

\section{Summary and Conclusions}
\label{sec:summary}

In this paper, we explore how statistical effects and sample selection affect the best-fit slope of the FP.  Since different coefficients are expected depending on the physical conditions and geometry very close to the black hole, and whether X-rays are dominated by  optically thin  jet synchrotron or inverse Compton, the FP can be exploited as a tool to diagnose radiative processes of accreting black holes.   Previous FP studies have been very insightful.  However, by using a more sophisticated (Bayesian) regression technique, it is possible to refine the FP even further.  Uncertainties on our best-fit coefficients are generally smaller than other FP studies, allowing more decisive statements on  radiative processes.   Also, our technique provides the first estimates of the intrinsic scatter about the FP measured directly from the data.

Ideally, one desires the largest possible dynamic range in mass and luminosity to obtain the most reliable FP coefficients.  We discuss how including only one class of AGN in FP studies severely limits the dynamic range and potentially decreases the reliability of the inferred coefficients.  However, we confirm the claims of \citetalias{falcke04} and \citetalias{koerding06} that jet X-rays  from the most massive \smbh s (i.e., $>$$\sim$10$^8~\msun$) are strongly affected by synchrotron cooling.   This fact makes it difficult to incorporate the most massive \smbh s in FP studies and then uniformly test if X-ray emission is dominated by the jet or by the corona.  For the most massive black holes, real X-ray data should be used to test corona models; however, if jet dominated, then ``X-ray luminosities'' must be extrapolated from lower-frequency wavebands (to be sure one is comparing only \textit{optically thin} jet synchrotron emission across the entire mass scale).

An advantage of our Bayesian regression analysis is that it is robust enough to measure reliable FP coefficients even if the most massive \smbh s are excluded.  Thus, we can test the dominate X-ray radiation mechanism without making any \textit{a priori} assumption on the radiation mechanism.  We find that a subset of sub-Eddington black holes with $\mbh <10^8~\msun$  (i.e., the contracted KFC subsample) favors optically thin jet synchrotron X-rays, provided their radio spectra are flat/inverted. Our analysis  excludes with high probability ($p>0.997$) the possibility that sub-Eddington black holes emit X-rays from a corona surrounding a RIAF with very low radiative efficiency (i.e., $q=2.3$), or surrounding a radiatively efficient $q=1$ accretion flow.  

We add a subset of  very massive accreting black holes with uncontroversially jet dominated SEDs (i.e., \bl\ objects) to our sample of $M<10^8~\msun$ sub-Eddington black holes.  For the reasons described above, the inclusion of these \bl\ objects does not provide new constraints on the X-ray radiation mechanism.  Rather, we use  these \bl\ objects to illustrate the systematic challenges to including the most massive black holes in FP regressions.  In particular, we show how the effects of synchrotron cooling from jet dominated SEDs are increasingly important in the X-ray waveband (and even in the optical waveband for FR~I galaxies and some LBLs) as black hole mass increases; we show that synchrotron cooling  will bias FP regressions if not taken into account.  We argue that one should model SEDs to measure the luminosity of their optically thin jet emission, although for HBLs one can alternatively extrapolate  optically thin synchrotron luminosities from the optical. 

The inclusion of \bl\ objects increases the dynamic range in black hole mass and luminosity,  allowing more accurate estimates of the FP coefficients especially when we only add HBLs to the lower-mass black hole sample.  For example, the uncertainties on our KFC+SDSS-HBL regression coefficients are $\sigma_{\xi_R} = \pm 0.04$, $\sigma_{\xi_M} = \pm 0.06$, compared to $\sigma_{\xi_R} = \pm 0.09$, $\sigma_{\xi_M} = \pm 0.09$ for the contracted KFC sample.  We find $\log L_x  = (1.45 \pm 0.04) \log L_R - (0.88 \pm 0.06) \log \mbh  - 6.07 \pm 1.10$, with $\sigma_{int}=0.07 \pm 0.05$~dex, and a similar regression is obtained if we debeam our \bl\ objects.  

  Our uniform SED modeling of a large number of \bl\ objects, combined with our adopted Bayesian regression technique, makes this one of the most accurate FP regressions yet performed. Note, the above correlation cannot be applied to higher accretion rate black holes, or to black holes without flat or inverted radio spectra.  These black holes may  instead  emit coronal X-rays and follow a different relation.  We also argue that, because of several observational effects due to relativistic beaming, inclusion of \bl\ objects allow more reliable coefficients than including FR~I galaxies, which may also have jet dominated SEDs.  On purely observational grounds, we advocate the removal of FR~I galaxies from FP studies unless  these difficulties are rigorously addressed.

A challenge to fitting the FP is that every object follows its own FP  in principle (i.e., depending on its particular values of $\alpha_R$, $p$, and $q$).  Also, accreting black holes are variable: their accretion rates, $\alpha_R$, $p$, and $q$ values, etc., change with time, meaning even individual sources are expected to follow different FP correlations over time.  Thus, statistical regressions only teach us about the ``average'' black hole.  Complicating the matter even more is that a magnetized corona may not be a totally distinct  entity from a collimated magnetized outflow at the base of a jet \citep[e.g.,][]{markoff05}, and observational signatures differentiating the two mechanisms are rather subtle.   In reality, both coronal and optically thin jet emission likely contribute to the observed X-ray emission, so the ``jet'' vs.\ ``corona'' models are best interpreted as  statements on which component is more dominant.  Furthermore, factors other than just luminosity are important, as evidenced, e.g., by XTE  J1550-564 showing different contributions of observed optically thin jet emission to the X-ray (at similar luminosities) depending on if the black hole is heading into outburst or declining back into quiescence \citep{russell10}.   

The above complications make using the FP to distinguish X-ray processes a challenging, but not impossible, exercise.    That inclusion of  accreting black holes at only certain accretion rates affects the intrinsic scatter and inferred coefficients  supports that GBH accretion states can indeed be mapped to different classes of AGN; thus, AGN unification is clearly more complicated than just orientation \citep[see, e.g.,][]{richards11}.    Looking toward the future, further exploration quantifying how parameter space beyond accretion rate, for example, environment, black hole spin, etc., can affect GBH-AGN mappings would be very exciting. 

\section*{Acknowledgments}
We thank the anonymous referee for insightful comments that  improved this manuscript.  We also thank Scott Trager for helpful discussions regarding measuring central black hole masses for our \bl\ objects.  R.M.P.\ and S.M.\ acknowledge support from a Netherlands Organization for Scientific Research (NWO) Vidi Fellowship.  S.M. also acknowledges support from The European CommunityÕs Seventh Framework Programme (FP7/2007-2013) under grant agreement number ITN 215212 ÒBlack Hole Universe.Ó  B.K.\  acknowledges support by NASA through Hubble Fellowship grant \#HF-51243.01 awarded by the Space Telescope Science Institute, which is operated by the Association of Universities for Research in Astronomy, Inc., for NASA, under contracted NAS 5-26555.  S.F.A.\ acknowledges support from NASA grant NNX09AF89G. 

\bibliographystyle{mn2e}
%\bibliography{../../../papers.bib}
%\begin{thebibliography}{98}
%\expandafter\ifx\csname natexlab\endcsname\relax\def\natexlab#1{#1}\fi

%\end{thebibliography}

%\appendix

\bsp

\label{lastpage}

\end{document}